\documentclass[preprint]{aastex}     

\usepackage{amsmath} 
\usepackage{graphicx}  
\usepackage{dcolumn}  
\usepackage{bm}        
\usepackage{lineno}   
\usepackage{hyperref}
\usepackage{xspace}

\shorttitle{Anisotropies in the Arrival Direction Distribution of Cosmic Rays}
\shortauthors{IceCube Collaboration}

\def \relInt   {$\Delta N/\langle N\rangle$\xspace}
\def \relIntI   {$\delta I = \Delta N/\langle N\rangle$\xspace}
 
\def \nside    {$N_\text{side}$\xspace}                  
\def \mus      {$\mu\text{s}$\xspace}                      
\def \muG      {$\mu\text{G}$\xspace}                    
\def \gRay     {$\gamma$-ray\xspace}                  
\def \datmap   {$N(\alpha,\delta)$\xspace}          
  
\def \dt       {$\Delta t$\xspace}                          
\def \alm      {$a_{\ell m}$\xspace}                   
         
\def \PolSpice {\texttt{PolSpice}\xspace}
\def \apriori  {\textsl{a priori}\xspace}
\def\eq#1{\begin{linenomath}\begin{equation}#1\end{equation}\end{linenomath}}

\begin{document}
  
\title{Observation of Anisotropy in the Arrival Directions of Galactic Cosmic Rays\\
at Multiple Angular Scales with IceCube}

\author{
IceCube Collaboration:
R.~Abbasi\altaffilmark{1},
Y.~Abdou\altaffilmark{2},
T.~Abu-Zayyad\altaffilmark{3},
J.~Adams\altaffilmark{4},
J.~A.~Aguilar\altaffilmark{1},
M.~Ahlers\altaffilmark{5},
D.~Altmann\altaffilmark{6},
K.~Andeen\altaffilmark{1},
J.~Auffenberg\altaffilmark{7},
X.~Bai\altaffilmark{8}
% ...
}
\altaffiltext{1}{Dept.~of Physics, University of Wisconsin, Madison, WI 53706, USA}
\altaffiltext{2}{Dept.~of Physics and Astronomy, University of Gent, B-9000 Gent, Belgium}
\altaffiltext{3}{Dept.~of Physics, University of Wisconsin, River Falls, WI 54022, USA}
\altaffiltext{4}{Dept.~of Physics and Astronomy, University of Canterbury, Private Bag 4800, Christchurch, New Zealand}
\altaffiltext{5}{Dept.~of Physics, University of Oxford, 1 Keble Road, Oxford OX1 3NP, UK}
\altaffiltext{6}{III. Physikalisches Institut, RWTH Aachen University, D-52056 Aachen, Germany}
\altaffiltext{7}{Dept.~of Physics, University of Wuppertal, D-42119 Wuppertal, Germany}
\altaffiltext{8}{Bartol Research Institute and Department of Physics and Astronomy, University of Delaware, Newark, DE 19716, USA}
% ...

\begin{abstract}

Between May 2009 and May 2010, the IceCube neutrino detector at the South Pole
recorded 32 billion muons generated in air showers produced by cosmic rays with
a median energy of 20 TeV.  With a data set of this size, it is possible to probe the 
southern sky for per-mille anisotropy on all angular scales in the arrival 
direction distribution of cosmic rays.  Applying a power spectrum analysis to the
relative intensity map of the cosmic ray flux in the southern hemisphere, we show that
the arrival direction distribution is not isotropic, but shows significant structure 
on several angular scales.  In addition to previously reported large-scale structure
in the form of a strong dipole and quadrupole, the data show small-scale structure 
on scales between $15^\circ$ and $30^\circ$.  The skymap exhibits several localized 
regions of significant excess and deficit in cosmic ray intensity.  The relative 
intensity of the smaller-scale structures is about a factor of 5 weaker than that 
of the dipole and quadrupole structure.  The most significant structure, an excess localized at 
(right ascension $\alpha=122.4^\circ$ and declination $\delta=-47.4^\circ$), extends over at least $20^\circ$ in right ascension 
and has a post-trials significance of $5.3\sigma$.  The origin of this anisotropy is still unknown.

\end{abstract}

\keywords{astroparticle physics --- cosmic rays}

\section{Introduction}\label{sec:Introduction}

The IceCube detector, deployed 1450\,m below the surface of the South Polar ice
sheet, is designed to detect upward-going neutrinos from astrophysical sources.  
However, it is also sensitive to downward-going muons produced in cosmic ray 
air showers.  To penetrate the ice and trigger the detector, the muons must 
possess an energy of at least several hundred GeV, which means they are produced 
by primary cosmic rays with energies in excess of several TeV.  Simulations show 
that the detected direction of 
an air shower muon is typically within $0.2^\circ$ of the direction of the
primary particle, so the arrival direction distribution of muons recorded in
the detector is also a map of the cosmic ray arrival directions between about 
1\,TeV and several 100\,TeV.  IceCube is currently the only instrument that can
produce such a skymap of cosmic ray arrival directions in the southern sky.  It
records several $10^{10}$ cosmic ray events per year, which makes it possible
to study anisotropy in the arrival direction distribution at the $10^{-4}$
level and below.

It is believed that charged cosmic rays at TeV energies are accelerated in 
supernova remnants in the Galaxy.  It is also expected that interactions of 
cosmic rays with Galactic magnetic fields should completely scramble their 
arrival directions.  For example, the Larmor radius of a proton with 10 TeV 
energy in a \muG magnetic field is approximately 0.01\,pc, orders of magnitude 
less than the distance to any potential accelerator.  Nevertheless, multiple
observations of anisotropy in the arrival direction distribution of
cosmic rays have been reported on large and small angular scales, mostly from
detectors in the northern hemisphere.  These deviations from isotropy in the
cosmic ray flux between several TeV and several hundred TeV are at the 
part-per-mille level, according to data from the Tibet AS$\gamma$
array~\citep{Amenomori:2005dy,Amenomori:2006bx}, the Super-Kamiokande
Detector~\citep{Guillian:2007}, the Milagro Gamma Ray
Observatory~\citep{Abdo:2008kr,Abdo:2008aw}, ARGO-YBJ~\citep{Vernetto:2009xm},
and EAS-TOP~\citep{Aglietta:2009mu}.
Recently, a study of muons observed with the IceCube detector has
revealed a large-scale anisotropy in the southern sky that is similar to that
detected in the north~\citep{Abbasi:2010mf}.  

In this paper, we present the results of a search for cosmic ray 
anisotropy on all scales in the southern sky with data recorded 
between May 2009 and May 2010 with the IceCube detector in its 
59-string configuration.  An angular power spectrum analysis 
reveals that the cosmic ray skymap as observed by IceCube is 
dominated by a strong dipole and quadrupole moment, but it also
exhibits significant structure on scales down to about $15^\circ$.  
This small-scale structure is about a factor 5 weaker in relative 
intensity than the dipole and quadrupole and becomes visible when 
these large-scale structures are subtracted from the data.  A 
comprehensive search for deviations of the cosmic ray flux from 
isotropy on all angular scales reveals several localized regions 
of cosmic ray excess and deficit, with a relative intensity of 
the order of $10^{-4}$.  The most significant structure 
is located at right ascension $\alpha=122.4^\circ$ and declination 
$\delta=-47.4^\circ$ and has a significance of $5.3\,\sigma$ after 
correcting for trials.  A comparison with data taken with fewer 
strings in the two years prior to this period confirms that these 
structures are a persistent feature of the southern sky.

The paper is organized as follows.  In this section, we give a short 
summary of previous observations, almost exclusively in the northern 
hemisphere, of anisotropy in the cosmic ray arrival skymap at TeV 
energies.  After the description of the IceCube detector and the data 
set used for this analysis (Section~\ref{sec:icecube}), the analysis 
techniques and results are presented in Section~\ref{sec:Analysis}.  
In Section~\ref{sec:systematics}, we show the outcome of several 
systematic checks of the analysis.  The results are summarized
and compared to Milagro results in the northern hemisphere in 
Section~\ref{sec:Conclusions}.

\subsection{Past Observations of Large- and Small-Scale Anisotropy}
\label{subsec:prevObs}

The presence of a large-scale anisotropy in the distribution of charged cosmic rays can be
caused by several effects.  For example, configurations of the heliospheric
magnetic field and other fields in the neighborhood of the solar system may be
responsible.  In this case, it is expected that the strength of the anisotropy
should weaken with energy due to the increasing magnetic rigidity of the
primary particles.  The present data cannot unambiguously
support or refute this hypothesis.  Measurements from the Tibet AS$\gamma$
experiment indicate that the anisotropy disappears above a few hundred
TeV~\citep{Amenomori:2006bx}, but a recent analysis of EAS-TOP data appears to
show an increase in the amplitude of the anisotropy above
400\,TeV~\citep{Aglietta:2009mu}.

Existing data sets have also been searched for a time-dependent modulation of
the anisotropy, which could be due to solar activity perhaps correlated with the eleven-year 
solar cycle.  Results are inconclusive at this point.  Whereas the Milagro data 
exhibit an increase in the mean depth of a large deficit region in the field of 
view over time~\citep{Abdo:2008aw}, no variation of the anisotropy with the
solar cycle has been observed in Tibet AS$\gamma$ data~\citep{Amenomori:2010yr}.
If these results are confirmed with more data recorded over longer time periods,
different structures might show a different long-term behavior.

A large-scale anisotropy can also be caused by any relative motion of the Earth through
the rest frame of the cosmic rays.  The intensity of the cosmic ray flux should be
enhanced in the direction of motion and reduced in the opposite direction,
causing a dipole anisotropy in the coordinate frame where the direction of
motion is fixed.  However, the Earth's motion through space is complex and a
superposition of several components, and the rest frame of the cosmic ray
plasma is not known.  If we assume the cosmic rays are at rest with respect to
the Galactic Center, then a dipole of amplitude 0.35\,\% should be observed due
to the solar orbit about the Galactic Center.  Such a dipole anisotropy, which
would be inclined at about $45^\circ$ with respect to the celestial equator,
was first proposed by \cite{Compton:1935}.  Although the effect is strong
enough to be measured by modern detectors, it has not been observed.  This null
result likely indicates that galactic cosmic rays co-rotate with the local Galactic
magnetic field \citep{Amenomori:2006bx}.  

The motion of the Earth around the Sun also causes a dipole in the arrival
directions of cosmic rays.  The dipole is aligned with the ecliptic plane, and
its strength is expected to be of order $10^{-4}$.  This solar dipole effect
has been observed by the Tibet AS$\gamma$ experiment \citep{Amenomori:2004bf}
and Milagro \citep{Abdo:2008aw} and provides a sensitivity test for all methods
looking for large-scale anisotropy in equatorial coordinates.

In addition to the large-scale anisotropy, data from several experiments in
the northern hemisphere indicate the presence of small-scale structures with scales 
of order $10^\circ$.  Using seven years of data, the Milagro collaboration
published the detection of two regions of enhanced flux with amplitude
$10^{-4}$ and a median energy of 1\,TeV with significance $>10\,\sigma$
\citep{Abdo:2008kr}.  The same excess regions also appear on skymaps produced
by ARGO-YBJ \citep{Vernetto:2009xm}.

Small-scale structures in the arrival direction distribution may indicate
nearby sources of cosmic rays, although the small Larmor radius at TeV energies
makes it impossible for these particles to point back to their sources unless
some unconventional propagation mechanism is assumed \citep{Malkov:2010yq}.
Diffusion from nearby supernova remnants, magnetic funneling
\citep{Drury:2008ns}, and cosmic ray acceleration from magnetic reconnection in
the solar magnetotail \citep{Lazarian:2010sq} have all been suggested as
possible causes for the small-scale structure in the northern hemisphere.

\subsection{Analysis Techniques}\label{subsec:paperStructure}

While the presence of large-scale structure in the southern sky has already
been established using IceCube data \citep{Abbasi:2010mf}, there has not been a
search of the southern sky for correlations on smaller angular scales.  In this
paper, we present a comprehensive study of the cosmic ray arrival directions in
IceCube which includes, but is not limited to, the search for small-scale
structures. 

Large and small-scale structure have traditionally been analyzed with very
different methods. The presence of a large-scale anisotropy is usually established 
by fitting the exposure-corrected arrival direction distribution in right ascension 
to the first few elements of a harmonic series~\citep{Amenomori:2006bx}.  While
essentially a one-dimensional method, the procedure can be applied to the right
ascension distribution in several declination bands to probe the strength of
dipole and quadrupole moments as a function of declination \citep{Abdo:2008aw}.
To search for small-scale structure, the estimation for an isotropic sky is
compared to the actual arrival direction distribution to find significant
deviations from isotropy \citep{Abdo:2008kr,Vernetto:2009xm}. 

Since both the large and small-scale structure in the cosmic ray data are
currently unexplained, it is not obvious whether a ``clean'' separation between
large and small scales is the right approach.  The anisotropy in the arrival 
direction distribution might be a superposition of several effects, with the 
small-scale structure being caused by a different mechanism than the large-scale 
structure, or it might be the result of a single mechanism producing a complex
skymap with structure on all scales.

The analysis presented in this paper makes use of a number of complementary
methods to study the arrival direction distribution without prior separation
into searches for large and small-scale structure.  The basis of this study 
is the angular power spectrum of the arrival direction distribution.  A power
spectrum analysis decomposes the skymap into spherical harmonics and provides
information on the angular scale of the anisotropy in the map.  The power spectrum
indicates which multipole moments $\ell=(0,1,2,\ldots)$ in the spherical
harmonic expansion contribute significantly to the observed arrival direction
distribution.  To produce a skymap of the contribution of the $\ell\geq3$ 
multipoles, the strong contributions from the dipole ($\ell=1$) and quadrupole 
($\ell=2$) have to be subtracted first.  The residual map can then be studied 
for structure on angular scales corresponding to $\ell\geq3$.  This is the 
first search for structure at these scales in the arrival direction distribution 
of TeV cosmic rays in the southern sky.

\section{The IceCube Detector}\label{sec:icecube}

IceCube is a $\mathrm{km}^{3}$-size neutrino detector frozen into the glacial
ice sheet at the geographic South Pole.  The ice serves as the detector medium.
High-energy neutrinos are detected by observing the Cherenkov radiation from
charged particles produced by neutrino interactions in the ice or in the
bedrock below the detector.  

The Cherenkov light is detected by an array of Digital Optical Modules (DOMs)
embedded in the ice.  Each DOM is a pressure-resistant glass sphere that
contains a 25~cm photomultiplier tube (PMT) \citep{Abbasi:2010vc} and
electronics which digitize, timestamp, and transmit signals to the data
acquisition system~\citep{Abbasi:2008ym}.  The IceCube array contains 5160 DOMs
deployed at depths between 1450\,m and 2450\,m below the surface of the ice
sheet.  The DOMs are attached to 86 vertical cables, or strings, which are used
for deployment and to transmit data to the surface.  The horizontal distance 
between strings in the standard detector geometry is about 125\,m, while the
typical vertical spacing between consecutive DOMs in each string is about 17 \,m. 
Six strings are arranged into a more compact configuration, with smaller spacing 
between DOMs, at the bottom of the detector, forming DeepCore, designed to extend 
the energy reach of IceCube to lower neutrino energies.  On the ice surface 
sits IceTop, an array of detectors dedicated to the study of the energy
spectrum and composition of cosmic rays with energies between 500\,TeV and 1\,EeV, 
several orders of magnitude larger in energy than the cosmic rays studied in this 
analysis.  All data used in this work comes from the IceCube in-ice detector only.

Construction of IceCube has recently been completed with all 86 strings deployed.  
The detector has been operating in various configurations since 
2005~\citep{Achterberg:2006md}.  Between 2007 and 2008, it operated with 22 
strings deployed (IC22), between 2008 and 2009 with 40 strings (IC40), and 
between 2009 and 2010 with 59 strings (IC59).

IceCube is sensitive to all neutrino flavors. Muon neutrinos,
identified by the ``track-like'' signature of the muon produced in a
charged-current interaction, form the dominant detection channel.  Muons
produced by astrophysical neutrinos are detected against an overwhelming
background of muons produced in cosmic ray air showers in the atmosphere above
the detector. IceCube searches are most sensitive to neutrino
sources in the northern hemisphere, where the Earth can be used as a filter
against atmospheric muons~\citep{Abbasi:2009iv}.

While atmospheric muons are a background for neutrino astrophysics,
they are a valuable tool in the analysis of the cosmic rays that produce them.
The downgoing muons preserve the direction of the cosmic ray air shower, and
thus the cosmic ray primary, and can be used to study the arrival
direction distribution of cosmic rays at energies above roughly 10\,TeV.

\subsection{DST Data Set}\label{subsec:dst}

The trigger rate of downgoing muons is about 0.5\,kHz in IC22, 1.1\,kHz in IC40,
and 1.7\,kHz in IC59.  This rate is of order $10^6$ times the neutrino rate,
and too large to allow for storage of the raw
data.  Instead, downgoing muon events are stored in a separate Data Storage and
Transfer (DST) format suitable for recording high-rate data at the South Pole.
The DST format is used to store the results of an online reconstruction
performed on all events that trigger the IceCube detector.  Most of the data
are downward-going muons produced by cosmic ray air showers.  Because of the
high trigger rate, the DST filter stream is used to save a very limited set of
information for every event. Basic event parameters such as energy estimators 
are stored, while digitized waveforms are only transmitted for a limited subset 
of events. 
The data are encoded in a compressed format that allows for the transfer 
of about $3$\,GB per day via the South Pole Archival and Data 
Exchange (SPADE) satellite communication system.

The main trigger used for physics analysis in IceCube is a simple majority
trigger which requires coincidence of 8 or more DOMs hit in the deep ice 
within a 5~\mus window. In order to pass the trigger condition, those hits 
have to be in coincidence with at least one other hit in the nearest or
next-to-nearest neighboring DOM within $\pm 1\,\mu s$ (local coincidence hits).  
Triggered events are reconstructed using two fast on-line
algorithms~\citep{Ahrens:2003fg}. The first reconstruction is a line-fit
algorithm based on an analytic $\chi^2$ minimization. It produces an
initial event track from the position and timing of the hits and the total
charge, but it does not account for the geometry of the Cherenkov cone and 
the scattering and absorption of photons in the ice. The second algorithm is 
a maximum likelihood-based muon track reconstruction,
seeded with the line-fit estimate of the arrival direction.  The likelihood
reconstruction is more accurate, but also more computationally expensive, so it
is applied only when at least ten optical sensors are triggered by the event.
The analysis presented in this work uses only events reconstructed with the
maximum likelihood algorithm.

In addition to particle arrival directions, the DST data also contain the
number of DOMs and hits participating in the event, as well as the total number
of triggered strings, and the position of the center of gravity of the event.   
The number of DOMs in the event can be used as a measure of the energy of the 
primary cosmic ray.  Above 1 TeV, the energy resolution is of order of
$0.5$ in $\Delta(\log(E))$, where $E$ is the energy of the primary cosmic ray.  
Most of the uncertainty originates in the physics of the air shower. 
In this energy range, we are dominated by air showers containing muons with
energies near the threshold necessary to reach the deep ice. Fluctuations in the 
generation of these muons are the main contribution 
to the uncertainty in the determination of the energy of the primary cosmic ray.

\subsection{Data Quality Cuts, Median Energy and Angular Resolution}\label{subsec:dataQuality}

The analysis presented in this paper uses the DST data collected 
 during IC59 operations between 2009 May 20 and 2010 May 30. 
The data set contains approximately $3.4\times10^{10}$ muon 
events detected with an integrated livetime of 334.5 days.  A cut in zenith 
angle to remove misreconstructed tracks near the horizon (see below) reduces 
the final data set to $3.2\times10^{10}$ events.

Simulated air showers are used to evaluate the median angular resolution of the
likelihood reconstruction and the median energy of the downgoing muon DST data 
set.  The simulated data are created using the standard air shower Monte Carlo 
program CORSIKA\footnote{COsmic Ray SImulations for KAscade:
\href{http://www-ik.fzk.de/corsika/}{\path{http://www-ik.fzk.de/corsika/}}}~\citep{Heck:1998a}.  
The cosmic ray spectrum and composition are simulated using the polygonato model of
\cite{Hoerandel:2002yg}, and the air showers are generated with the SIBYLL
model of high-energy hadronic interactions~\citep{Ahn:2009wx}.

\begin{figure}[t]
  \begin{center}
     $\begin{array}{cc}
      \includegraphics[width=.45\textwidth]{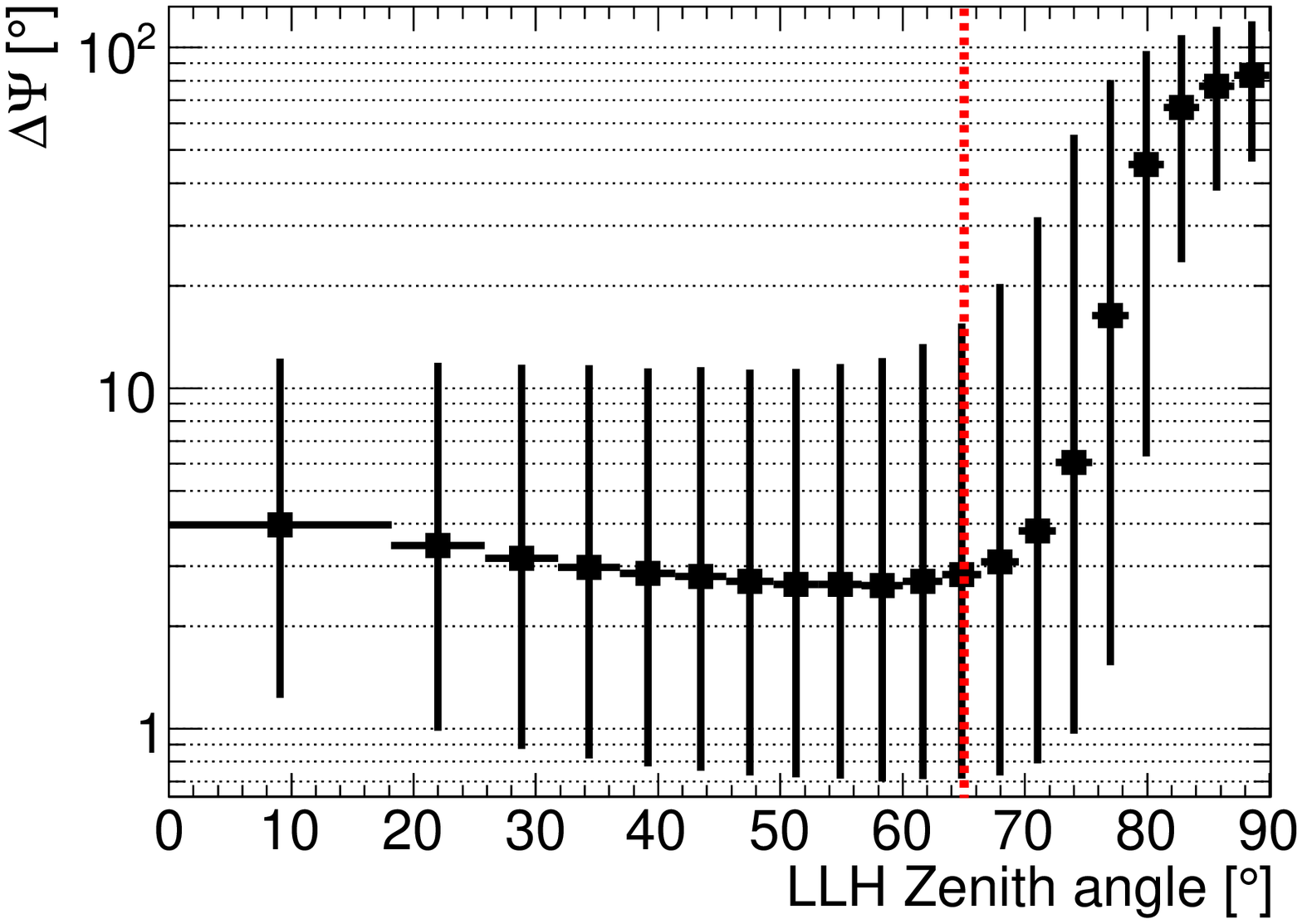} &
     \includegraphics[width=.45\textwidth]{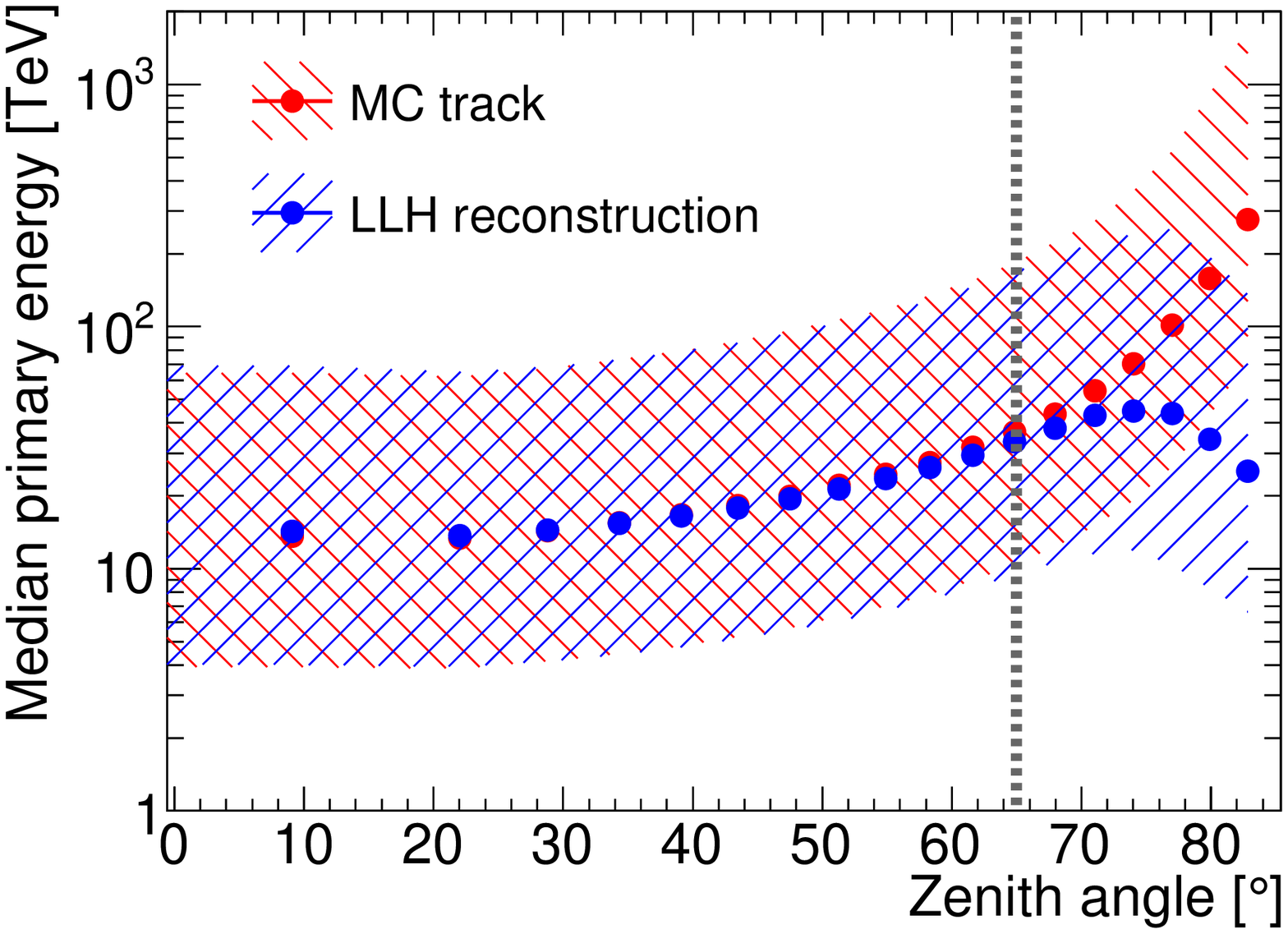}  
  \end{array}$
  \end{center}
  \caption{Median angular resolution {\em (left)} and median energy {\em (right)} 
  as a function of zenith angle for simulated cosmic ray events.  The error 
  bars on the left plot and the hatched regions on the right one correspond 
  to a $68\%$ containing interval.  The median primary energy is shown both as a function
  of the true zenith angle (MC track) and the reconstructed zenith angle
  (LLH reconstruction), while the median angular resolution {\em (left)} is 
  shown as a function of the reconstructed zenith angle only.  The dotted vertical line at $\theta = 65^\circ$ 
  indicates the cut in zenith angle performed in this work. }
  \label{f:zenenergy}
\end{figure}

The simulations show that, for zenith angles smaller than $65^\circ$,
the median angular resolution is $3^\circ$.  
This is not to be confused with the angular resolution of IceCube
for neutrino-induced tracks (better than $1^\circ$), where more sophisticated 
reconstruction algorithms and more stringent quality cuts are applied. The resolution 
depends on the zenith angle of the muon.  Fig.~\ref{f:zenenergy} (left) shows the 
median angular resolution as a function of zenith angle.  
The resolution improves from $4^\circ$ at small zenith angles to about $2.5^\circ$ near $60^\circ$.  
The larger space angle error at small zenith angles is caused by the detector geometry, 
which makes it difficult to reconstruct the azimuth angle for near-vertical showers.   
Consequently, with the azimuth angle being essentially unknown, the angular error can be large.  
For zenith angles greater than $65^\circ$, the angular resolution degrades markedly.
The reason is that more and more events with apparent zenith angle 
greater than $65^\circ$ are misreconstructed 
tracks of smaller zenith angle and lower energy.  The energy threshold for 
muon triggers increases rapidly with slant depth in the atmosphere and ice, 
and the statistics at large zenith angle become quite poor.  We restrict our 
analysis to events with zenith angles smaller than $65^\circ$.  Within this 
range, the angular resolution is roughly constant and much smaller than the
angular size of arrival direction structure we are trying to study.

Using simulated data, we estimate that the overall median energy of the primary
cosmic rays that trigger the IceCube detector is 20\,TeV. Simulations show that at this energy
the detector is more sensitive to protons than to heavy nuclei like iron.  The median energy
increases monotonically with the true zenith angle of the primary particle
(Fig.~\ref{f:zenenergy} (right)) due to the attenuation of low-energy muons with
increasing slant depth of the atmosphere and ice.  The median energy also
increases as a function of reconstructed zenith angle. Near the horizon, the large fraction 
of misreconstructed events causes the median energy to fall.

\section{Analysis}\label{sec:Analysis}

The arrival direction distribution of cosmic rays observed by detectors 
like IceCube is not isotropic.  Nonuniform exposure to different parts 
of the sky, gaps in the uptime, and other detector-related effects will cause
a spurious anisotropy in the measured arrival direction distribution even 
if the true cosmic ray flux is isotropic.  Consequently, in any search for 
anisotropy in the cosmic ray flux, these detector-related effects need to be
accounted for.  The first step in this search is therefore the creation of 
a ``reference map'' to which the actual data map is compared.  The reference 
map essentially shows what the skymap would look like if the cosmic ray flux 
was isotropic.  It is not in itself isotropic, because it includes the detector 
effects mentioned above.  The reference map must be subtracted from the real 
skymap in order to find regions where the actual cosmic ray flux deviates from 
the isotropic expectation.

In this section, we first describe the construction of the reference map 
for the subsequent analysis.  The reference map is then compared to the 
actual data map, and a map of the relative cosmic ray intensity is produced.
We then perform several analyses to search for the presence 
of significant anisotropy in the relative intensity map.

\subsection{Calculation of the Reference Level}\label{subsec:background}

For the construction of a reference map that represents the detector
response to an isotropic sky, it is necessary to determine the exposure
of the detector as a function of time and integrate it over the livetime.
We use the method of \cite{Alexandreas:1993} to calculate the exposure
from real data.  This technique is commonly used in \gRay astronomy to 
search for an excess of events above the exposure-weighted isotropic 
reference level.

The method works as follows.  The sky is binned into a fine grid in 
equatorial coordinates (right ascension $\alpha$, declination $\delta$).  
Two sky maps are then produced.  The data map \datmap stores the arrival 
directions of all detected events.  For each detected event that is stored 
in the data map, 20 ``fake'' events are generated by keeping the local zenith 
and azimuth angles $(\theta,\phi)$ fixed and calculating new values for 
right ascension using times randomly selected from 
within a pre-defined time window \dt bracketing the time of the event
being considered.  These fake events are stored in the reference map
with a weight of 1/20.  Using 20 fake events per real event, the 
statistical error on the reference level can be kept small.

Created in this way, the events in the reference map have the same local 
arrival direction distribution
as the real events.  Furthermore, this ``time scrambling'' method naturally 
compensates for variations in the event rate, including the presence of gaps 
in the detector uptime.  The buffer length \dt needs to be chosen such that 
the detector conditions remain stable within this period.  Due to its unique 
location at the South Pole, the angular acceptance of IceCube is stable over 
long periods.  The longest \dt used in this analysis is one day, and the 
detector stability over this time period has been verified by $\chi^2$-tests
comparing the arrival direction distributions at various times inside the
window.  The IceCube detector is, in fact, stable over periods longer than 
24 hours.

Deviations from isotropy are known to bias estimates of the reference level
produced by this method.  In the vicinity
of a strong excess, the method can create artificial deficits, as the events
from the excess region are included in the estimation of the reference level.
Similarly, there can be artificial excesses near strong deficits.
In searches for point sources, the effect is usually negligible, but it can
become significant in the presence of extended regions of strong excess or
deficit flux.

Since the Earth rotates by $15^\circ$ every hour, the right ascension range 
of the scrambled data is $15^\circ/\mathrm{hour} \times\Delta t$, so any 
structure in the data map that is larger than 
$15^\circ/\mathrm{hour}\times\Delta t$ will also appear in the reference map
and therefore be suppressed in the relative intensity map \relInt.  
For example, $\Delta t=2~\text{hr}$ will suppress structures larger than 
$30^\circ$ in the relative intensity map.  To be sensitive to large-scale 
structure such as a dipole, a time window of 24 hours (or higher) must be 
used.

\begin{figure}[t]
  \begin{center}	
    \includegraphics[width=0.495\textwidth]{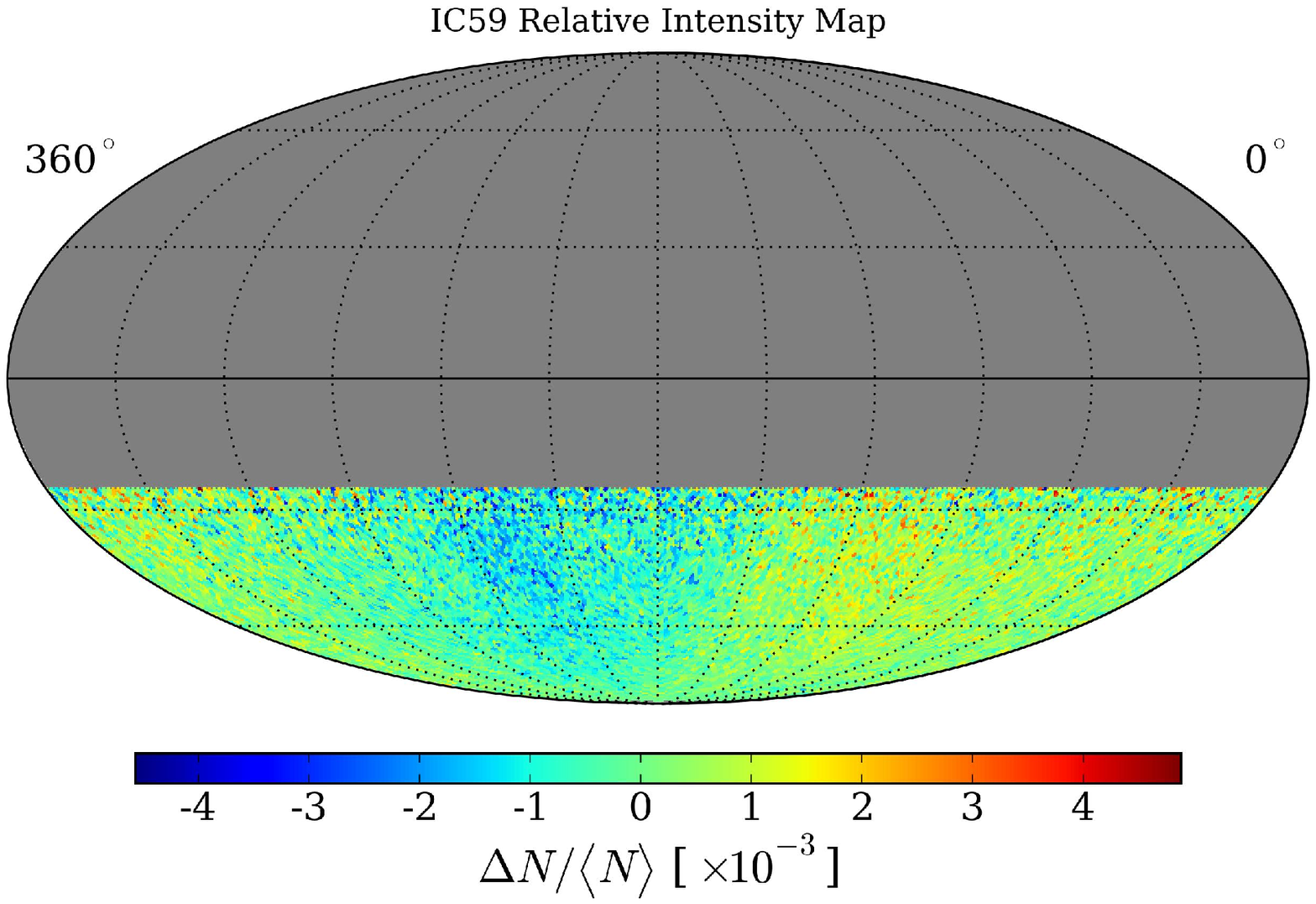}
    \includegraphics[width=0.495\textwidth]{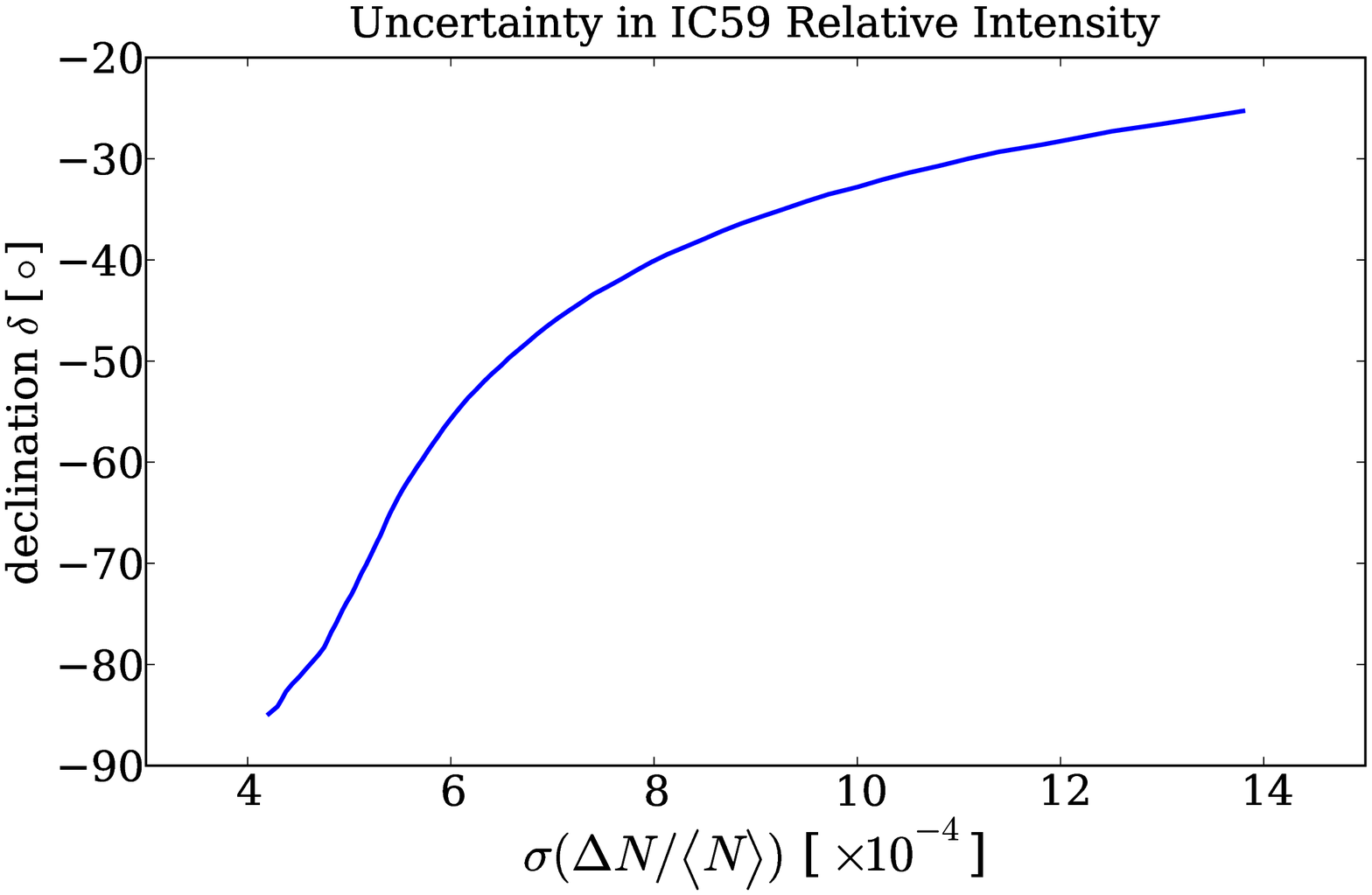}
    \caption{{\em Left:} Relative intensity \relInt of the IC59 data in equatorial 
    coordinates, produced with a time window of 24 hours.  {\em Right:} Dependence of the statistical error on the declination.}
    \label{fig:ic59RelInt}
  \end{center}
\end{figure}

\subsection{Relative Intensity and Significance Maps}\label{subsec:maps}

Once the data and reference maps are calculated, deviations from isotropy can
be analyzed by calculating the relative intensity
\begin{linenomath}
  \begin{equation}\label{eq:relInt}
    \frac{\Delta N_i}{\langle N\rangle_i} =
      \frac{N_i(\alpha,\delta)-\langle N_i(\alpha,\delta)\rangle}{\langle N_i(\alpha,\delta)\rangle}.
  \end{equation}
\end{linenomath}
which gives the amplitude of deviations from the isotropic expectation 
in each angular bin $i$.  The deviations from isotropy can also be
expressed in terms of a statistical significance using the method of
\cite{LiMa:1983}.  We report both relative intensity maps and significance maps
in this paper.

\begin{figure}[t]
\begin{center}	
$\begin{array}{cc}	
  \includegraphics[width=.495\textwidth]{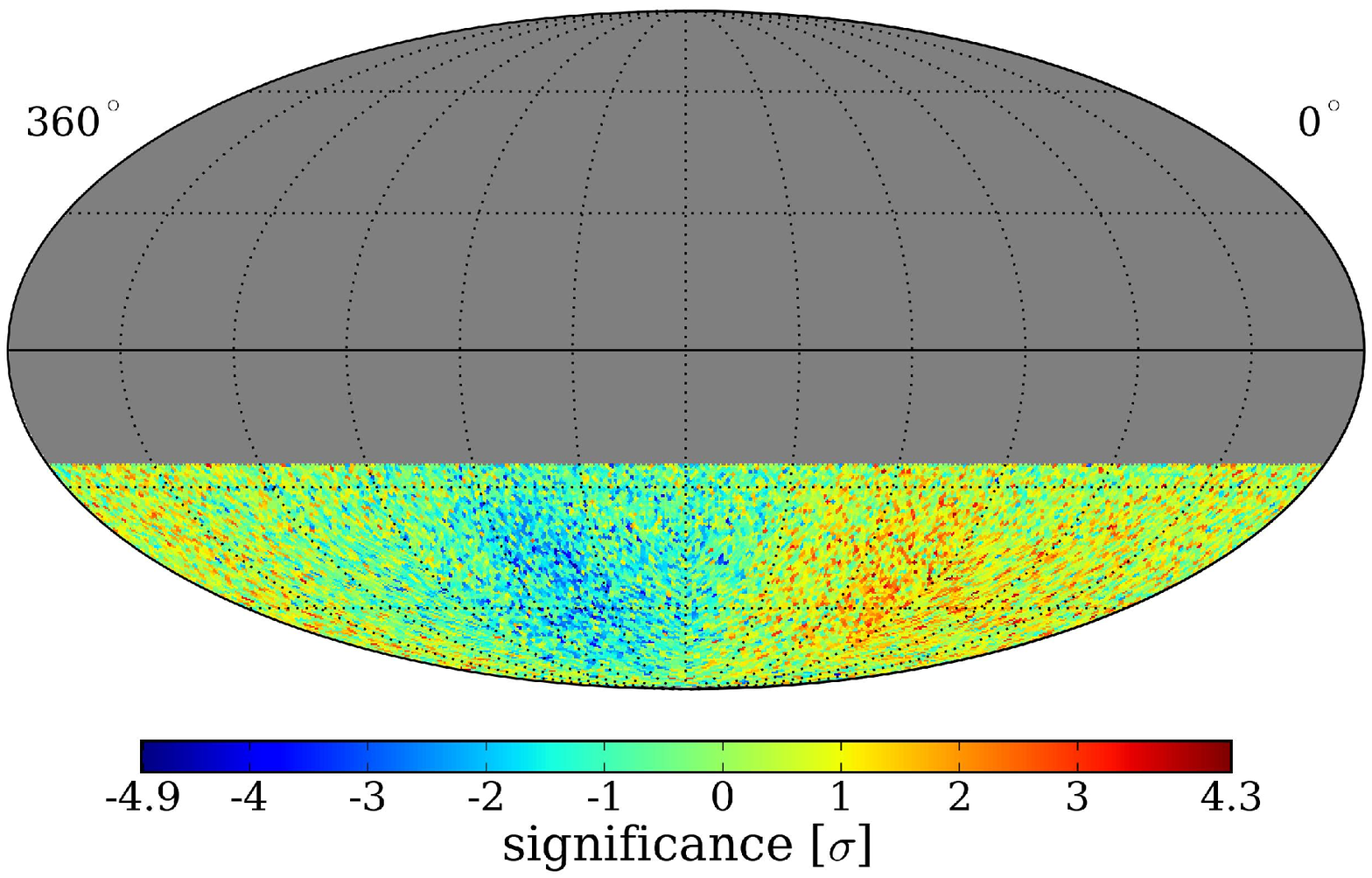}
   \includegraphics[width=.495\textwidth]{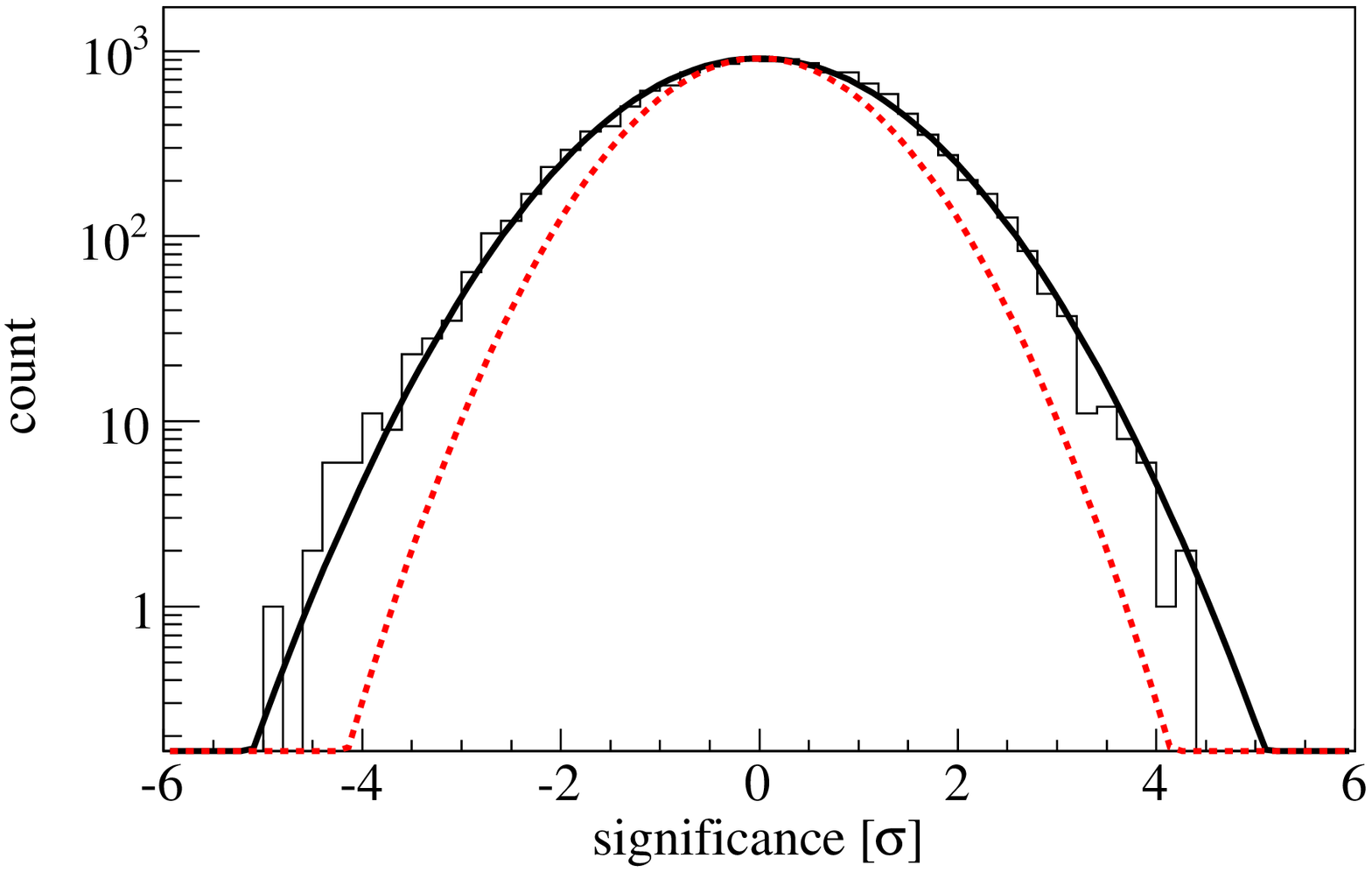}
\end{array}$
\end{center}
	\caption{{\em Left}: Significance sky map of the IC59 data in equatorial 
  coordinates, produced using a time window of 24 hours. {\em Right}:
  1d-distribution of the significance values together with the best-fit 
  (black solid line) performed with a Gaussian function. For comparison, a Gaussian
  function of mean zero and unit variance (red dashed line), expected from an
  isotropic sky, has been superimposed.}
	\label{fig:largeScaleMap}
\end{figure}

The analyses in this paper use the HEALPix\footnote{Hierarchical 
Equal-Area isoLatitude Pixelization of the sphere:
\href{http://healpix.jpl.nasa.gov/}{\path{http://healpix.jpl.nasa.gov}}.}
library for the production of skymaps~\citep{Gorski:2004by}.  HEALPix produces an
equal-area division of the unit sphere with pixels of roughly equal shape.  The
resolution of the HEALPix grid is defined by a parameter called \nside, which
is related to the number of pixels in the grid by
$N_\text{pix}=12N_\text{side}^2$.  Here, \nside$=64$ has been
chosen, so that the sky is divided into 49\,152 pixels with an average pixel size 
of about $0.9^\circ$.
Due to the zenith angle cut of $65^\circ$ discussed in
Section~\ref{subsec:dataQuality}, the pixels above declination $\delta=-25^\circ$
are masked in the analysis.  This leaves 14\,196 pixels in the region between
$\delta=-25^\circ$ and the celestial South Pole at $\delta=-90^\circ$.  The sky
maps are plotted in equatorial coordinates using an equal-area homolographic
projection.

Figure~\ref{fig:ic59RelInt} (left) shows the relative intensity when a 24-hour time
window is used to estimate the reference level.  The map exhibits clear structures.
The most obvious features are a broad excess in
the relative counts near right ascension $105^\circ$, and a broad deficit 
near $225^\circ$.  The relative intensity in the excess (and
deficit) region is of order $10^{-3}$.  This structure is the large-scale
anisotropy first observed in the analysis of the IC22 data set and reported
in~\cite{Abbasi:2010mf}.  Since the IC59 data set is larger than the IC22 data
set by an order of magnitude, it is now possible to see the large scale
structure directly in the data without further rebinning or averaging over many
pixels.

Figure~\ref{fig:ic59RelInt} (right) shows the statistical error on the relative intensity map.
Relative intensity skymaps have declination-dependent statistical uncertainties
due to the fact that the detector acceptance decreases with larger zenith
angle.  Since IceCube is located at the South Pole, the relative intensity
exhibits large fluctuations near the horizon, corresponding to declinations
$\delta>-30^\circ$ .  Such edge effects are not as severe for skymaps of the
significance of the fluctuations, though one must note that the location of
structures with large (or small) significance may not coincide with regions 
of large (or small) relative intensity.

Fig.~\ref{fig:largeScaleMap} (left) shows the significance map corresponding to
the relative intensity map shown in Figure~\ref{fig:ic59RelInt}.  The right
panel also shows a distribution of the significance values in each bin.  In an
isotropic skymap, the distribution of the significance values should be normal
(red dashed line).  However, the best Gaussian fit to the distribution (black solid line)
exhibits large deviations from a normal distribution caused by the large-scale 
structure.

\subsection{Angular Power Spectrum Analysis}\label{subsec:powerSpectrum}

To observe correlations between pixels at several angular scales,
we calculate the angular power spectrum of the relative intensity map
\relIntI described in Section~\ref{subsec:maps}.  The relative 
intensity can be treated as a scalar field which we expand in terms 
of a spherical harmonic basis,
\eq{\label{eq:sphHarmonics}
  {\delta I(\mathbf{\mathbf{u}}_i)} =
    \sum_{\ell=1}^{\infty}
    \sum_{m=-\ell}^{\ell} a_{\ell m} Y_{\ell m}(\mathbf{u}_i)
}
\eq{\label{eq:alm}
  a_{\ell m}
  \sim \Omega_{p} \sum_{i=0}^{N_{\mathrm{pix}}} \delta I(\mathbf{u}_{i}) Y_{\ell m}^{*}(\mathbf{u}_{i})~~.
}
In Eqs.~\eqref{eq:sphHarmonics} and \eqref{eq:alm}, the $Y_{\ell m}$ are the
Laplace spherical harmonics, the \alm are the multipole coefficients of the
expansion, $\Omega_{p}$ is the solid angle observed by each pixel (which is 
constant across the sphere in HEALPix), $\mathbf{u}_i=(\alpha_i,\delta_i)$ 
is the pointing vector associated with the $i^{\mathrm{th}}$ pixel, and 
$N_{\mathrm{pix}}$ is the total number of pixels in the skymap. The power 
spectrum for the relative intensity field is defined as the variance of the 
multipole coefficients \alm,
\eq{
  {\cal C}_{\ell} = \frac{1}{2 \ell + 1} \sum_{m=-\ell}^{\ell} | a_{\ell m} |^{2}~~.
   \label{eq:cldef}
}
The amplitude of the power spectrum at some multipole order $\ell$ is
associated with the presence of structures in the sky at angular scales of
about $180^\circ/\ell$.  In the case of complete and uniform sky coverage, a
straightforward Fourier decomposition of the relative intensity maps would
yield an unbiased estimate of the power spectrum. However, due to the limited
exposure of the detector, we only have direct access to the so-called
pseudo-power spectrum, which is the convolution of the real underlying power
spectrum and the power spectrum of the relative exposure map of the detector in equatorial coordinates.
In the case of partial sky coverage, the standard $Y_{\ell m}$ spherical harmonics 
do not form an orthonormal basis that we can use to expand the relative intensity 
field directly.  As a consequence of this, the pseudo-power spectrum displays a 
systematic correlation between different $\ell$ modes that needs to be corrected 
for.

The deconvolution of pseudo-power spectra has been a longstanding problem in
CMB astronomy, and there are several computationally efficient tools available 
from the CMB community.  (For a discussion on the bias introduced by partial sky 
coverage in power spectrum estimation and a description of several bias removal 
methods, see \cite{Ansari:2009ys}.)  To calculate the power spectrum of the IC59 data, 
we use the publicly available \PolSpice\footnote{\PolSpice website:
\href{http://prof.planck.fr/article141.html}{\path{http://prof.planck.fr/article141.html}}.}
software package \citep{Szapudi:2000xj, Chon:2003gx}.
  
In \PolSpice, the correction for partial sky bias is performed not on the power 
spectrum itself, but on the two-point correlation function of the relative intensity 
map. The two-point correlation function $\xi(\eta)$ is defined as
 \eq{
  {\xi (\eta)} = \langle \delta I(\mathbf{u}_{i}) \; \delta I(\mathbf{u}_{j}) \rangle~~,
  }
where $\delta I(\mathbf{u}_{k})$ is the observed relative intensity in the direction 
of the $k^{\mathrm{th}}$ pixel.  Note that $\xi(\eta)$ depends only on the angle 
$\eta$ between any two pixels. The two-point correlation function can be expanded 
into a Legendre series, 
\eq{
  {\xi (\eta)} = \frac{1}{4 \pi} \sum_{\ell = 0}^{\infty} (2 \ell + 1)\ {\cal C}_{\ell}\ P_{\ell}(\cos \eta)~~,
   \label{eq:xi}
 } 
where the ${\cal C}_{\ell}$ are the coefficients of the angular power spectrum and 
the $P_{\ell}$ are the Legendre polynomials.  The inverse operation
 \eq{
 {\cal C}_{\ell} = 2 \pi \int^{1}_{-1} \xi(\eta)\ P_{\ell}(\cos \eta)\ d(\cos \eta)
 \label{eq:xi2cl}
 }
can be used to calculate the angular power spectrum coefficients from a known two-point 
correlation function.  

In order to obtain an unbiased estimator of the true power spectrum, 
\PolSpice first calculates the \alm coefficients of both the relative 
intensity map and the relative exposure map doing a spherical harmonics 
expansion equivalent to that shown in Eq.\eqref{eq:alm}.  Pseudo-power 
spectra for both maps are computed from these coefficients using 
Eq.\eqref{eq:cldef}, and these spectra are subsequently converted 
into correlation functions using Eq.\eqref{eq:xi}.
An unbiased estimator ${\tilde{\xi}}(\eta)$ of the true correlation 
function of the data is computed by taking the ratio of the correlation 
functions of the relative intensity map and the relative exposure map.  
An estimate ${\cal \tilde{C}}_{\ell}$ of the true power spectrum can 
then be obtained from the corrected two-point correlation function 
using the integral expression shown in Eq.\eqref{eq:xi2cl}.

This process reduces the correlation between different $\ell$ modes 
introduced by the partial sky coverage.  Minor ringing artifacts associated with
the limited angular range over which the correlation function is evaluated are 
minimized by applying an apodization function to the correlation function 
in $\eta$-space as described in \cite{Chon:2003gx}.  The cosine apodization 
scheme provided by \PolSpice and used in this work allows the correlation 
function to fall slowly to zero at large angular scales where statistics are 
low, minimizing any ringing artifacts that could arise from the calculation of 
the power spectrum from the corrected correlation function using Eq.\eqref{eq:xi2cl}.

\begin{figure}[t]
  \begin{center}	
    \includegraphics[width=0.7\textwidth]{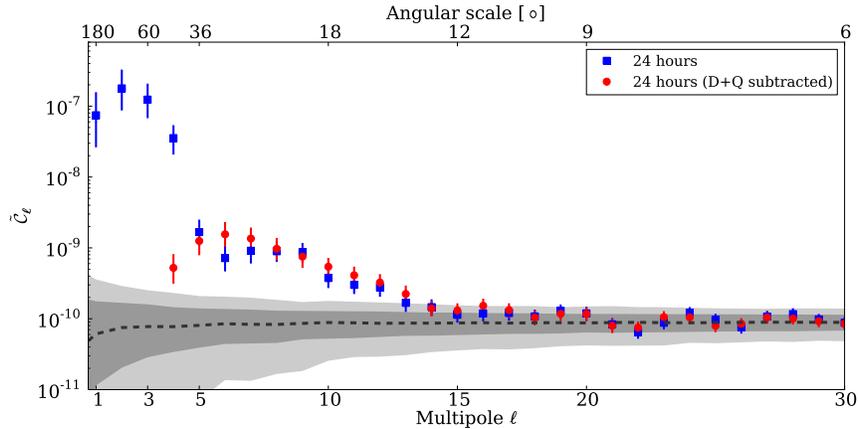}
    \caption{Angular power spectra for the relative intensity map shown in
    Fig.\,\ref{fig:ic59RelInt}.  The blue and red points show the power 
    spectrum before and after the subtraction of the dominant dipole and 
    quadrupole terms from the relative intensity map.  Errors bars are
    statistical, but a possible systematic error is discussed in the text.  
    The gray bands indicate the distribution of the power spectra in a large 
    sample of isotropic data sets, showing the 68\% (dark) and 95\% (light)
spread in the ${\cal \tilde{C}}_{\ell}$. }
    \label{fig:powerspectrum_all}
  \end{center}
\end{figure}

Fig.\,\ref{fig:powerspectrum_all} (blue points) shows the angular power
spectrum for the IC59 relative intensity map from Fig.\,\ref{fig:ic59RelInt}.
In addition to a strong dipole and quadrupole moment ($\ell=1,2$), higher order
terms up to $\ell=12$ contribute significantly to the skymap.  The error bars
on the ${\cal \tilde{C}}_{\ell}$ are statistical.  The gray bands indicate the 68\% 
and 95\% spread in the ${\cal \tilde{C}}_{\ell}$ for a large number of power spectra 
for isotropic data sets (generated by introducing Poisson fluctuations in the reference skymap).
As the ${\cal \tilde{C}}_{\ell}$ are still not entirely independent (even after
the correction for partial sky coverage is performed), a strong dipole moment
in the data can lead to significant higher order multipoles, and it is
important to study whether the structure for $3\leq\ell\leq12$ is a systematic
effect caused by the strong lower order moments $\ell=1,2$.
Fig.\,\ref{fig:powerspectrum_all} (red points) shows the angular power spectrum
after the strong dipole and quadrupole moments are removed from the relative
intensity map by a fit procedure described in the next section.  The plot
illustrates that after the removal of the lower order multipoles, indicated by
the drop in ${\cal \tilde{C}}_{\ell}$ for $\ell=1,2$ (both are consistent with
0 after the subtraction), most of the higher order terms are still present.
Only the strength of ${\cal \tilde{C}}_{3}$ and ${\cal \tilde{C}}_{4}$ is
considerably reduced (the former to a value that is below the range of the
plot). 

Regarding systematic uncertainties, for $\ell=3$ and $\ell=4$ the effects of
the strong dipole and quadrupole suggest that there is significant coupling
between the low-$\ell$ modes.  Therefore, we cannot rule out that ${\cal
\tilde{C}}_3$ and ${\cal \tilde{C}}_4$ are entirely caused by systematic
effects.  For the higher multipoles, the systematic effects of this distortion
are much lower.  After explicit subtraction of the $\ell=1$ and $\ell=2$ terms
the residual power spectrum agrees with
the original power spectrum within the statistical uncertainties.  Therefore,
we conclude that the systematic uncertainties in these data points are, at
most, of the same order as the statistical uncertainties.

In summary, the skymap of cosmic ray arrival directions contains significant
structures on scales down to $\sim 15^\circ$.  In the next sections, we
describe analysis techniques to make the smaller scale structure visible in the
presence of the much stronger dipole and quadrupole moments.  

\subsection{Subtraction of the Dipole and Quadrupole Moments}
\label{subsec:subtraction}

A straightforward approach to understand the contribution of higher order
multipoles and the corresponding structure in the skymap is to remove the
strong dipole and quadrupole moments from the relative intensity map and study
the residuals.  This requires a dipole and quadrupole fit to the IC59 map.
Once fit, the dipole and quadrupole can be subtracted from the skymap.
We fit the relative intensity map using the function
\begin{linenomath}
  \begin{multline}\label{eq:dqfit}
    \delta I(\alpha,\delta) =
    m_0 + p_x\cos{\delta}\cos{\alpha}
    + p_y\cos{\delta}\sin{\alpha} + p_z\sin{\delta} \\
      + \frac{1}{2}Q_1(3\cos^2{\delta}-1)
      + Q_2 \sin{2\delta}\cos{\alpha}
      + Q_3 \sin{2\delta}\sin{\alpha}
      + Q_4 \cos^2{\delta}\cos{2\alpha}
      + Q_5 \cos^2{\delta}\sin{2\alpha}.
  \end{multline}
\end{linenomath}
Equation~\eqref{eq:dqfit} is a multipole expansion of the relative count
distribution in terms of real-valued spherical harmonic functions, and follows
a normalization convention commonly used in CMB physics~\citep{Smoot:1979}.
The quantity $m_0$ is the ``monopole'' moment of the distribution, and
corresponds to a constant offset of the data from zero.  The values
$(p_x,p_y,p_z)$ are the components of the dipole moment, and the quantities
$(Q_1,\ldots,Q_5)$ are the five independent components of the quadrupole
moment.

\begin{figure}[t]
  \begin{center}
    \includegraphics*[width=0.7\textwidth,clip]{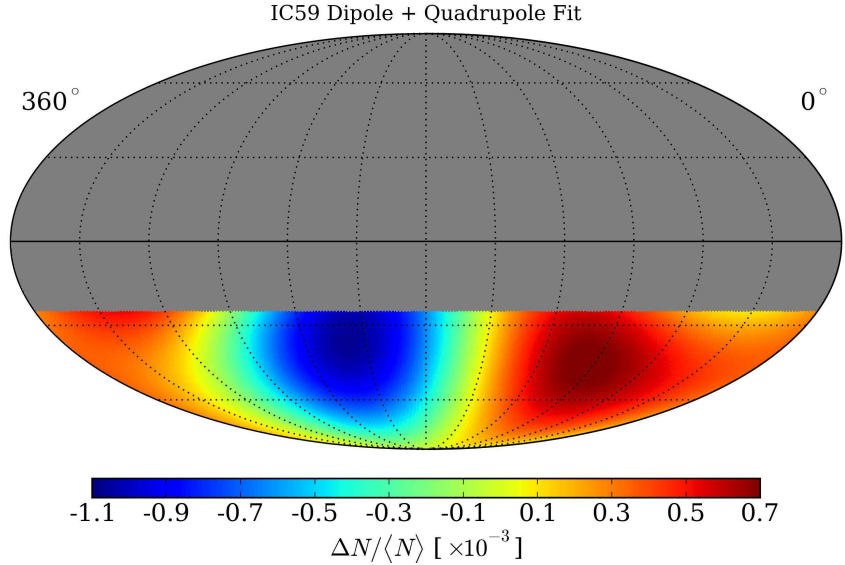}
    \caption{\label{fig:ic59_fitdq}
    Fit of Eq.~\eqref{eq:dqfit} to the IC59 relative intensity distribution
    \relInt shown in Fig.~\ref{fig:ic59RelInt}.}
  \end{center}
\end{figure}

\begin{table}[t]
  \begin{center}
  {\small
     \begin{tabular}{crrrrrrrrrr} \hline
      Coefficient  & \multicolumn{1}{c}{Value (stat. + syst.)}
                   & \multicolumn{9}{c}{Correlation Coefficients} \\
      $\quad$      & \multicolumn{1}{c}{($\times10^{-4}$)}
                   & \multicolumn{9}{c}{$\chi^2/\text{ndf}=14743/14187:~\text{Pr}(\chi^2|\text{ndf})=5.5\times10^{-4}$} \\
                   \hline
    $m_0$ & $ 0.32\pm2.26\pm0.28$ &  1.00 &        &        &        &        &
    &        &        &       \\
    $p_x$ & $ 2.44\pm0.71\pm0.30$ &  0.00 &  1.00 &        &        &        &
    &        &        &       \\
    $p_y$ & $-3.86\pm0.71\pm0.94$ &  0.00 &  0.00 &  1.00 &        &        &
    &        &        &       \\
    $p_z$ & $ 0.55\pm3.87\pm0.45$ &  1.00 &  0.00 &  0.00 &  1.00 &        &
    &        &        &       \\
    $Q_1$ & $ 0.23\pm1.70\pm0.17$ &  0.99 &  0.00 &  0.00 &  0.99 &  1.00 &
    &        &        &       \\
    $Q_2$ & $-2.95\pm0.49\pm0.74$ &  0.00 &  0.98 &  0.00 &  0.00 &  0.00 &
    1.00 &        &        &       \\
    $Q_3$ & $-8.80\pm0.49\pm0.50$ &  0.00 &  0.00 &  0.98 &  0.00 &  0.00 &
     0.00 &  1.00 &        &       \\
    $Q_4$ & $-2.15\pm0.20\pm0.50$ &  0.00 &  0.00 &  0.00 &  0.00 &  0.00 &
    0.00 &  0.00 &  1.00 &       \\
    $Q_5$ & $-5.27\pm0.20\pm0.06$ &  0.00 &  0.00 &  0.00 &  0.00 &  0.00 &
    0.00 &  0.00 &  0.00 &  1.00\\
    \hline
    \end{tabular}
  }

    \caption{\label{table:dqfit} Coefficients for the fit of Eq.~\eqref{eq:dqfit} to the IC59 relative intensity
    distribution.  The correlation coefficients indicate that there is some degeneracy between the contributions 
    of $p_x$ and $Q_2$, $p_y$ and $Q_3$, and $p_z$ and $Q_1$ due to the fact that the IceCube detector only has 
    a partial view of the sky.  The systematic error on the fit parameters is estimated using the results of a fit
    using anti-sidereal time as described in Sec.~\ref{subsec:antiSidereal}.}
  \end{center}
\end{table}

\begin{figure}[ht]
  \begin{center}
    \includegraphics*[width=0.495\textwidth,clip]{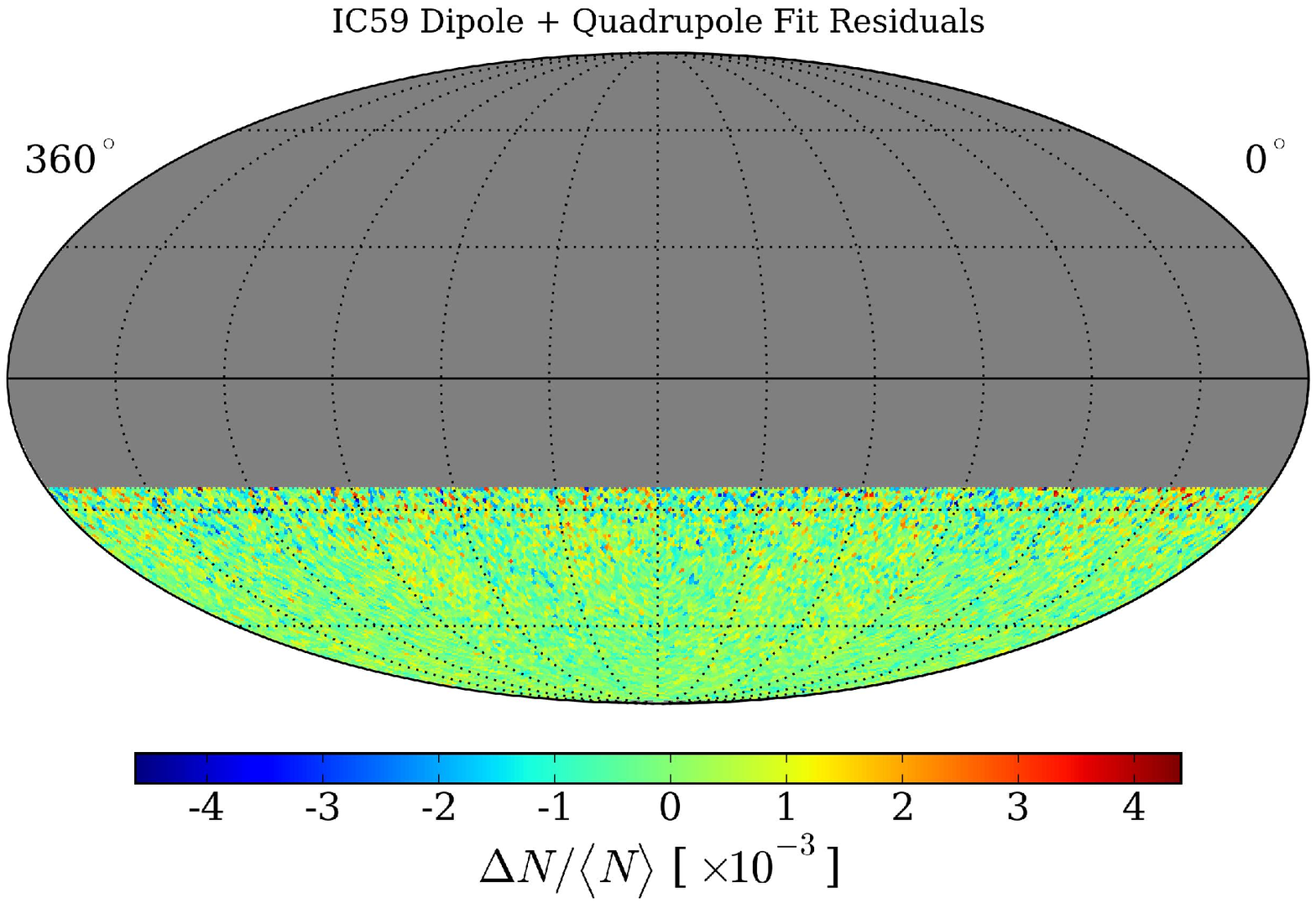}
    \includegraphics*[width=0.495\textwidth,clip]{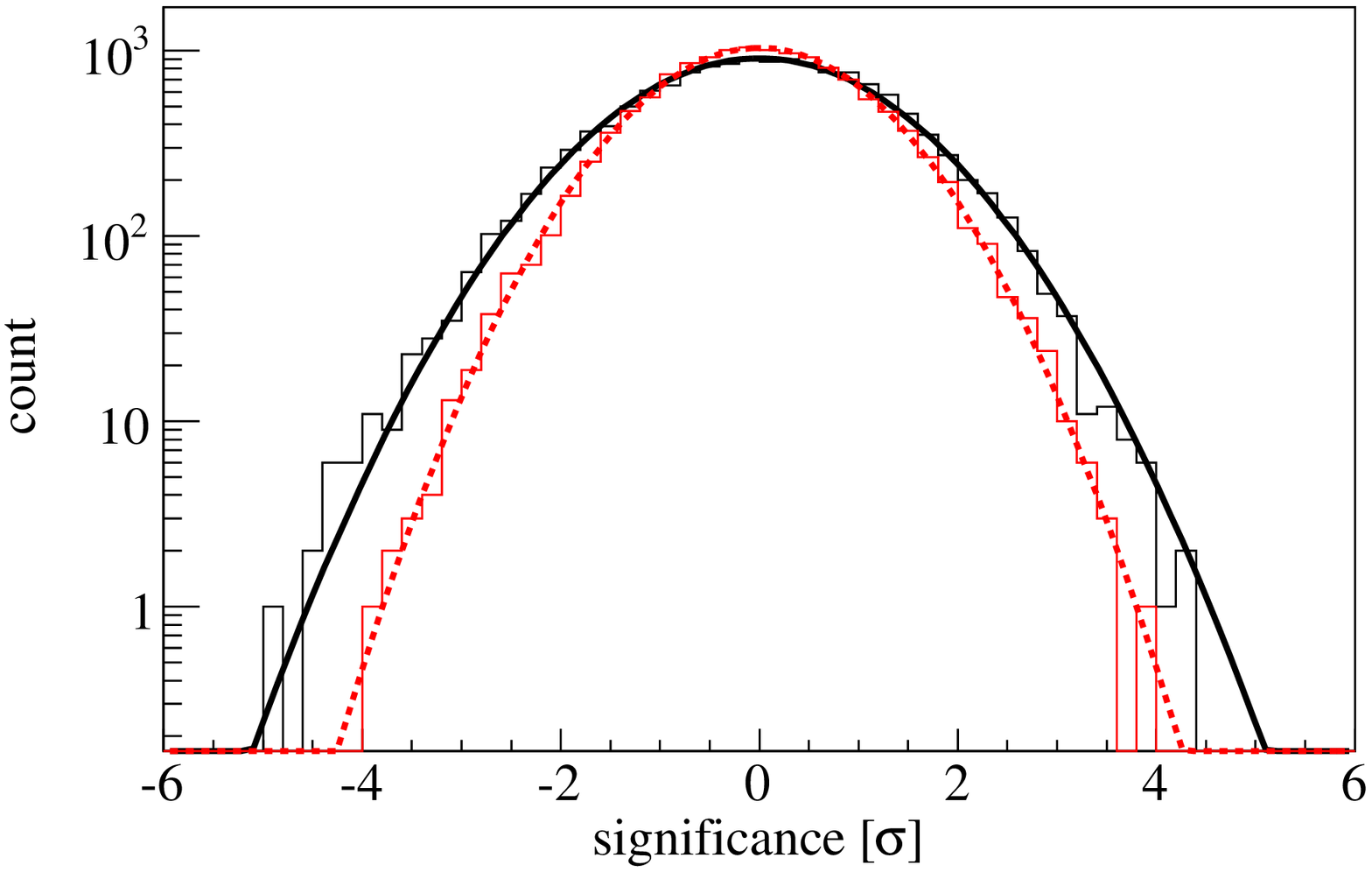}
  \caption{\label{fig:ic59FitdqRes} 
    {\em Left}: Residual of the fit of Eq.~\eqref{eq:dqfit} to the relative
    intensity distribution shown in Fig.~\ref{fig:ic59RelInt}.
    {\em Right:} Distribution of pixel significance values in the skymap 
    before (solid black line) and after (dashed red line) subtraction of 
    the dipole and quadrupole. Gaussian fits to the data yield a mean
    of $(-0.20\pm1.05)\times10^{-2}$ and a width of $1.23\pm0.01$ before the
    dipole and quadrupole subtraction, and $(0.28\pm0.89)\times10^{-2}$ and 
    $1.02\pm0.01$ after.
    }
  \end{center}
\end{figure}

The two-dimensional harmonic expansion of Eq.~\eqref{eq:dqfit} was fit to the
14\,196 pixels in the IC59 relative intensity map that lie between the
celestial South Pole and declination $\delta=-25^\circ$.  The best-fit dipole
and quadrupole coefficients are provided in Table~\ref{table:dqfit}, and the
corresponding sky distribution is shown in Fig.~\ref{fig:ic59_fitdq}.  By
themselves, the dipole and quadrupole terms can account for much of the
amplitude of the part-per-mille anisotropy observed in the IceCube data.  We
note that the quadrupole moment is actually the dominant term in the expansion,
with a total amplitude that is about $2.5$ times larger than the dipole
magnitude.  
However, the $\chi^2/\text{ndf}=14743/14187$ corresponds to a
$\chi^2$-probability of approximately $0.05\%$, so while the dipole and quadrupole are
dominant terms in the arrival direction anisotropy, they do not appear to
be sufficient to explain all of the structures observed in the angular
distribution of \relInt.  This result is consistent with the result of the angular power
spectrum analysis in Sec.~\ref{subsec:powerSpectrum}, which also indicates the
need for higher-order multipole moments to describe the structures in the
relative intensity skymap.

\begin{figure}[t]
  \begin{center}
  $\begin{array}{cc}
    \includegraphics*[width=0.495\textwidth,clip]{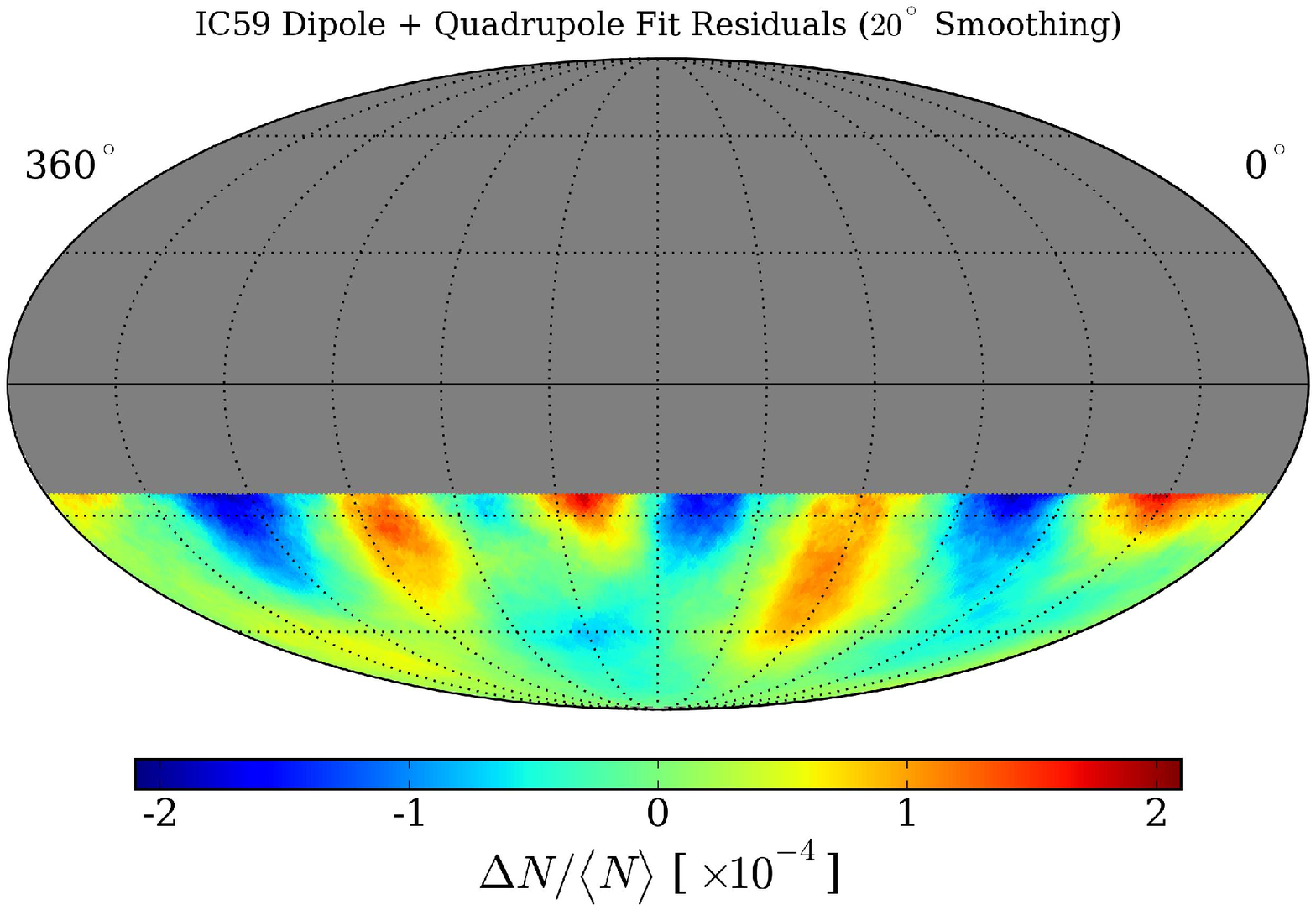}
    \includegraphics*[width=0.495\textwidth,clip]{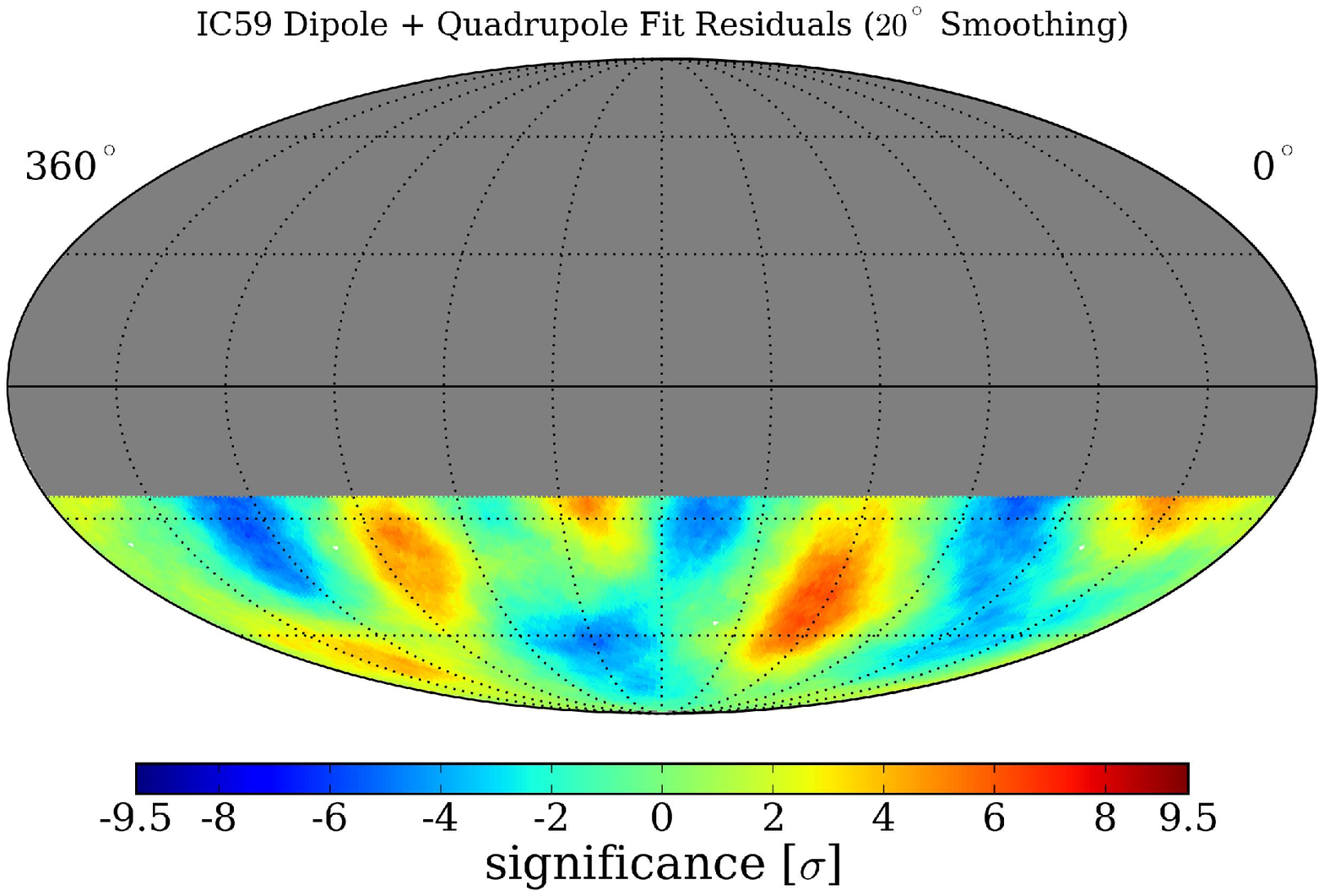}
  \end{array}$
  \caption{\label{fig:ic59FitdqResSmoothed}
    {\em Left}: Residual intensity map plotted with $20^\circ$
    smoothing.
    {\em Right}: Significances of the residual map (pre-trials), plotted with
    $20^\circ$ smoothing.}
  \end{center}
\end{figure}

Subtraction of the dipole and quadrupole fits from the relative intensity map
shown in Fig.~\ref{fig:ic59RelInt} yields the residual map shown in
Fig.~\ref{fig:ic59FitdqRes}.  The fit residuals are relatively featureless at
first glance, and the significance values are
well-described by a normal distribution, which is expected when no anisotropy
is present.  However, the bin size in this plot is not optimized for a study of
significant anisotropy at angular scales larger than the angular resolution 
of the detector.  To improve the sensitivity to larger features, we apply a 
smoothing procedure which simply takes the reference level and residual data counts 
in each bin and adds the counts from pixels within some angular radius of the 
bin.  This procedure results in a map with Poisson uncertainties, though the 
bins are no longer statistically independent.

The actual size of any possible excess or deficit region (and thus the optimal
smoothing scale) is not known \apriori.  Furthermore, the skymap may contain
several significant structures of different size, with the optimal smoothing
radius differing for each structure.  To make the search as comprehensive as
possible, we study the skymap on all smoothing scales from $3^\circ$ (the angular
resolution) to
$45^\circ$ in steps of $1^\circ$ and search for regions of high significance at
any location.  Applying this procedure, the two most significant localized excesses
on the sky are a region with a peak significance of $7.0\sigma$ at a smoothing
radius of $22^\circ$ at $(\alpha=122.4^\circ, \delta=-47.4^\circ)$, and a region
of peak significance $6.7\sigma$ at a smoothing radius of $13^\circ$ at
$(\alpha=263.0^\circ, \delta=-44.1^\circ)$.  These values do not account for
statistical trials due to the scan over smoothing radii or the scan for the
peak significance in the 14\,196 pixels. We have estimated the trial factors 
by applying the same search strategy to a large number of simulated isotropic data sets.
After trial factors are applied, 
the maximum significance of the ``hot spot'' with an optimal smoothing radius of  $22^\circ$ is reduced to 
$5.3\sigma$, and the ``hot spot'' at $13^\circ$ is reduced to $4.9\sigma$.

Skymaps of the relative intensity and the significance of the residual data are
plotted in Fig.\,\ref{fig:ic59FitdqResSmoothed}, where a smoothing radius of
$20^\circ$ has been used.  The radius is not optimal for any of the most
significant excesses, but with this choice all of the significant features
can be seen with reasonable resolution.  Compared to the intensity of the
dipole and quadrupole shown in Fig.\,\ref{fig:ic59RelInt}, the smaller
structures are weaker by about a factor of 5.

Table\,\ref{table:5sigmaspots} contains the location and optimal smoothing
scales of all the regions in the IC59 skymap which have a pre-trials
significance beyond $\pm5\sigma$.  The data also exhibit additional regions of
excess and deficit.  It is possible that the deficits are at least in part
artifacts of the reference level estimation procedure, which can produce artificial
deficits around regions of significant excess counts (or in principle, excesses
in the presence of strong physical deficits).  While several of the deficit and
excess regions are observed at large zenith angles near the edge of the IC59
exposure region, we do not believe these features are statistical fluctuations
or edge effects.  As we will show in Section~\ref{subsec:ic22ic40comp}, the
features are also present in IC22 and IC40 data, and grow in significance as
the statistics increase.

\begin{figure}[t]
  \begin{center}
  $\begin{array}{cc}
    \includegraphics*[width=0.499\textwidth,clip]{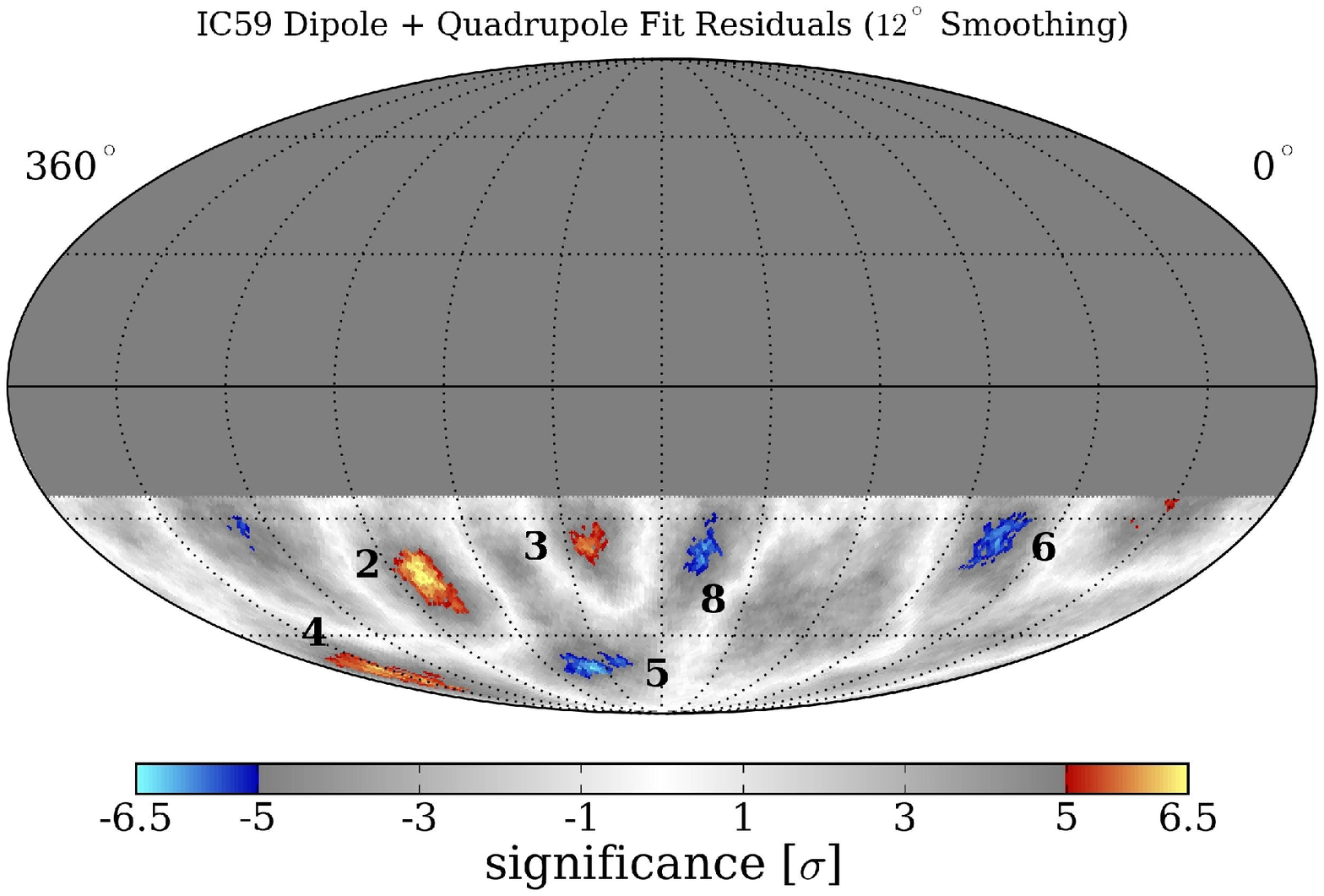}
    \includegraphics*[width=0.499\textwidth,clip]{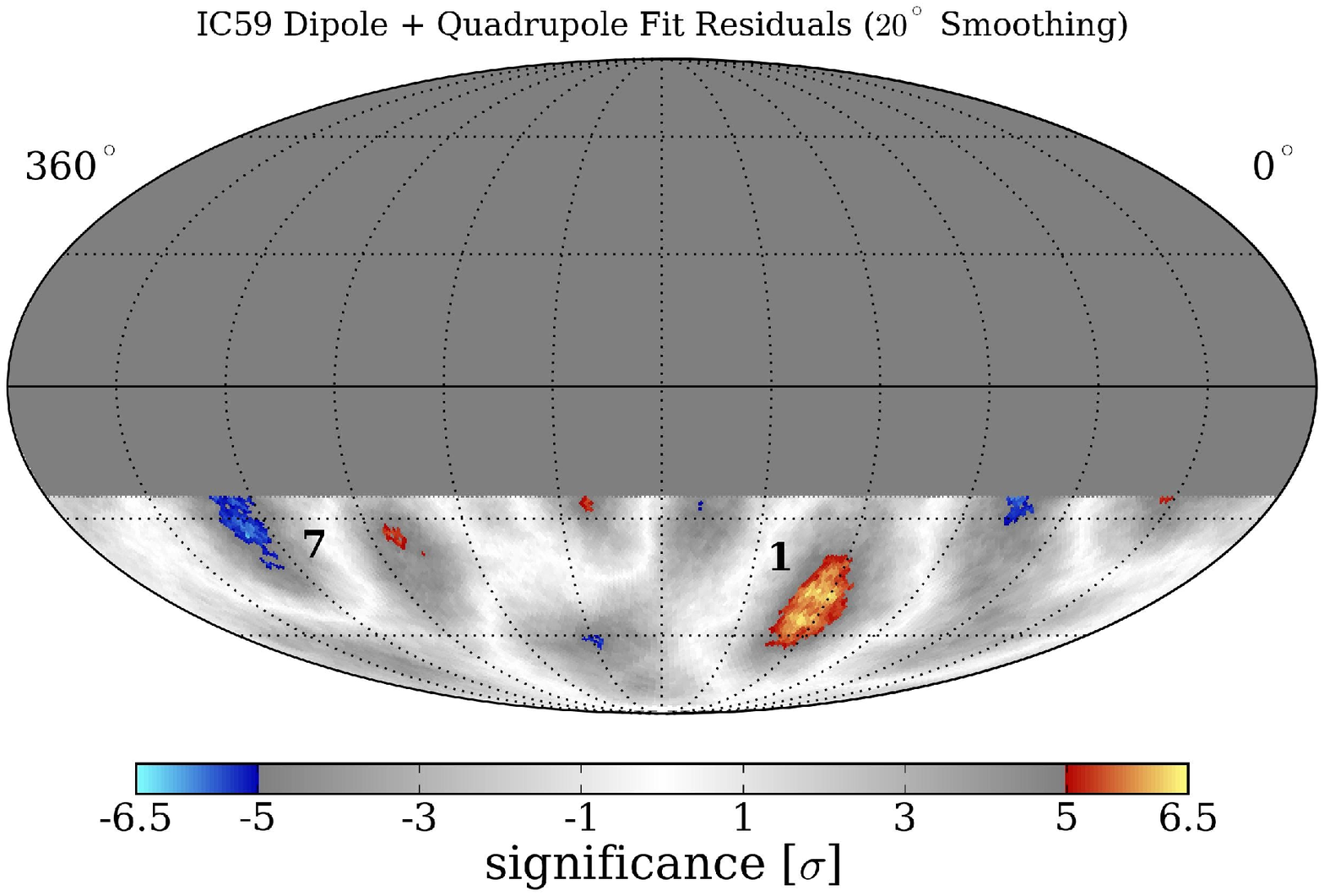}
  \end{array}$
  \caption{\label{fig:ic59FitdqResThresholds}
    {\em Left}: Significances of the IC59 residual map plotted with $12^\circ$
    smoothing.
    {\em Right}: Significances of the IC59 residual map plotted with
    $20^\circ$ smoothing.  The regions with a pre-trial significance 
    larger than $\pm5\sigma$ are 
    indicated according to the numbers used in Table~\ref{table:5sigmaspots}.}
  \end{center}
\end{figure}

Fig.\,\ref{fig:ic59FitdqResThresholds} shows the significance maps with regions 
with a pre-trial significance larger than $\pm5\sigma$ indicated according to 
the numbers used in Table~\ref{table:5sigmaspots}.  Since the optimal scales vary
from region to region and no single smoothing scale shows all regions, we
show the maps with two smoothing scales, $12^\circ$ (left), and $20^\circ$ (right).

The angular power spectrum of the residual map is shown in red
in Fig.\,\ref{fig:powerspectrum_all}.  As expected, there is no significant
dipole or quadrupole moment left in the skymap, and the $\ell=3$ and $\ell=4$
moments have also disappeared or have been weakened substantially.  However,
the moments corresponding to $5\leq\ell\leq12$ are still present at the same
strength as before the subtraction, and indicate the presence of structure
of angular size $15^\circ$ to $35^\circ$ in the data.  The excesses and deficits
in Fig.~\ref{fig:ic59FitdqResSmoothed} correspond in size to these moments.

\begin{table}[ht]
  \begin{center}
  {\small
    \renewcommand{\arraystretch}{1.2}
    \begin{tabular}{cccccc} \hline
      region  & right ascension & declination & optimal scale & peak significance & post-trials\\
      \hline
      1 & $(122.4^{+4.1}_{-4.7})^\circ$ & $(-47.4^{+7.5}_{-3.2})^\circ$ &
      $22^\circ$ & $\ \ 7.0\sigma$ & $\ \ 5.3\sigma$ \\
      2 & $(263.0^{+3.7}_{-3.8})^\circ$ & $(-44.1^{+5.3}_{-5.1})^\circ$ &
      $13^\circ$ & $\ \ 6.7\sigma$ & $\ \ 4.9\sigma$ \\
      3 & $(201.6^{+6.0}_{-1.1})^\circ$ & $(-37.0^{+2.2}_{-1.9})^\circ$ &
      $11^\circ$ & $\ \ 6.3\sigma$ & $\ \ 4.4\sigma$ \\
      4 & $(332.4^{+9.5}_{-7.1})^\circ$ & $(-70.0^{+4.2}_{-7.6})^\circ$ &
      $12^\circ$ & $\ \ 6.2\sigma$ & $\ \ 4.2\sigma$ \\
      \hline
      5 & $(217.7^{+10.2}_{-7.8})^\circ$ & $(-70.0^{+3.6}_{-2.3})^\circ$ &
      $12^\circ$ & $-6.4\sigma$ & $-4.5\sigma$ \\
      6 & $(77.6^{+3.9}_{-8.4})^\circ$ & $(-31.9^{+3.2}_{-8.6})^\circ$ &
      $13^\circ$ & $-6.1\sigma$ & $-4.1\sigma$ \\
      7 & $(308.2^{+4.8}_{-7.7})^\circ$ & $(-34.5^{+9.6}_{-6.9})^\circ$ &
      $20^\circ$ & $-6.1\sigma$ & $-4.1\sigma$ \\
      8 & $(166.5^{+4.5}_{-5.7})^\circ$ & $(-37.2^{+5.0}_{-5.7})^\circ$ &
      $12^\circ$ & $-6.0\sigma$ & $-4.0\sigma$ \\
      \hline
    \end{tabular}
    \renewcommand{\arraystretch}{1}
  }
  \caption{\label{table:5sigmaspots}
    Location and optimal smoothing scale for regions of the IC59 skymap with 
    a pre-trials significance larger than $\pm5\sigma$.  The errors on the 
    equatorial coordinates indicate the range over which the significance
    drops by $1\,\sigma$ from the local extremum.}
  \end{center}
\end{table}

\clearpage

\subsection{A Filter for Structure on Small Angular Scales}
\label{subsec:timewindows}

In previous works~\citep{Abdo:2008kr,Vernetto:2009xm}, a different method is
applied to filter the lower $\ell$ terms and create skymaps showing the
small-scale structure.  In these analyses, the dipole and quadrupole moments
are not fit and subtracted, but suppressed by varying the time window
$\Delta t$ over which the reference level is estimated ({\it i.e.}, the length of
time in which the time scrambling, or any other method for generating an 
isotropic sky, is performed).  We apply this method to the IC59 data to compare
the results to the dipole and quadrupole subtraction outlined in
Sec.\,\ref{subsec:subtraction}.

Different time windows probe the presence of anisotropy at different angular
scales.  The time scrambling fits structures that are larger than
$15^{\circ}/\mathrm{hour} \times \Delta t$, and the angular size of a multipole
of order $\ell$ in the sky is $\sim 180^{\circ}/\ell$.  This implies that the
technique filters out modes with $\ell < 12~\mathrm{hours}/ \Delta t$
and reduces the magnitude of the modes near this threshold.  

\begin{figure}[t]
  \begin{center}
    \includegraphics[width=0.7\textwidth]{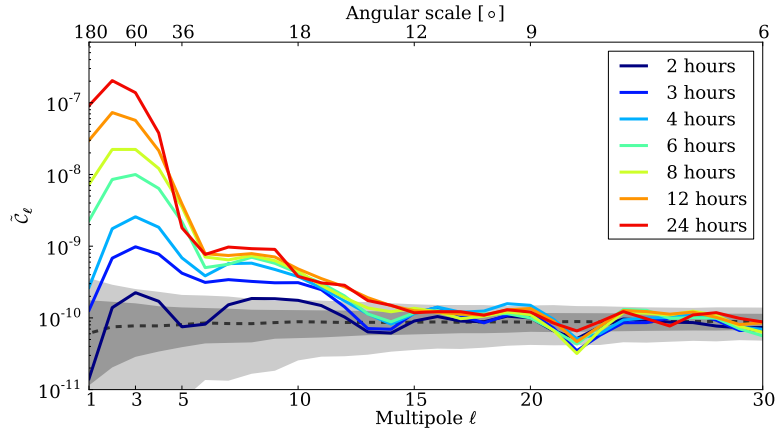}
    \caption{Power spectra for different values of the time scrambling
    period $\Delta t$. The filtering effect of the time scrambling on 
    large-scale structure can be easily seen as a monotonic reduction in the 
    strength of low-$\ell$ components of the power spectrum.  The grey bands
    show 1 and 2 sigma bands for a large set of isotropic skymaps.  See
    Fig.~\ref{fig:powerspectrum_all} and Section~\ref{subsec:powerSpectrum} for
    statistical uncertainties and a discussion of systematic uncertainties.}
    \label{fig:inttimes_ps}
  \end{center}
\end{figure}

The efficiency of the method in suppressing larger structures (low-$\ell$
moments) is demonstrated in Fig.\,\ref{fig:inttimes_ps}, where the angular
power spectra are plotted for relative intensity maps constructed with seven
values of $\Delta t$ between 2 hours and 24 hours.  As expected, the strength
of the low-order multipoles decreases monotonically with $\Delta t$.  However,
the power spectrum also reveals that the low-$\ell$ moments, in particular the
quadrupole term, are not completely removed from the data unless $\Delta t$ is
as small as 3 hours.  In addition, the choice of $\Delta t\leq3$\,hours also
appears to weaken the power observed in the modes $3\leq\ell\leq12$.
Consequently, the residual map from Sec.\,\ref{subsec:subtraction} and the
skymaps produced by choosing a small $\Delta t$ cannot be expected to agree in
all details.  Nevertheless, a comparison of the skymaps produced with the two
methods provides an important crosscheck.

To best compare this analysis to the results of
Section~\ref{subsec:subtraction}, the reference level is calculated using a
scrambling time window of $\Delta t=4$\,hours.  This choice of $\Delta t$ is
motivated by the angular power spectrum in Fig.\,\ref{fig:inttimes_ps}.  With
$\Delta t=4$\,hours, the spectrum shows the strongest suppression of the dipole
and quadrupole while still retaining most of power in the higher multipole
moments.  

\begin{figure}[t]
\begin{center}	
$\begin{array}{cc}	
   \includegraphics[width=.45\textwidth]{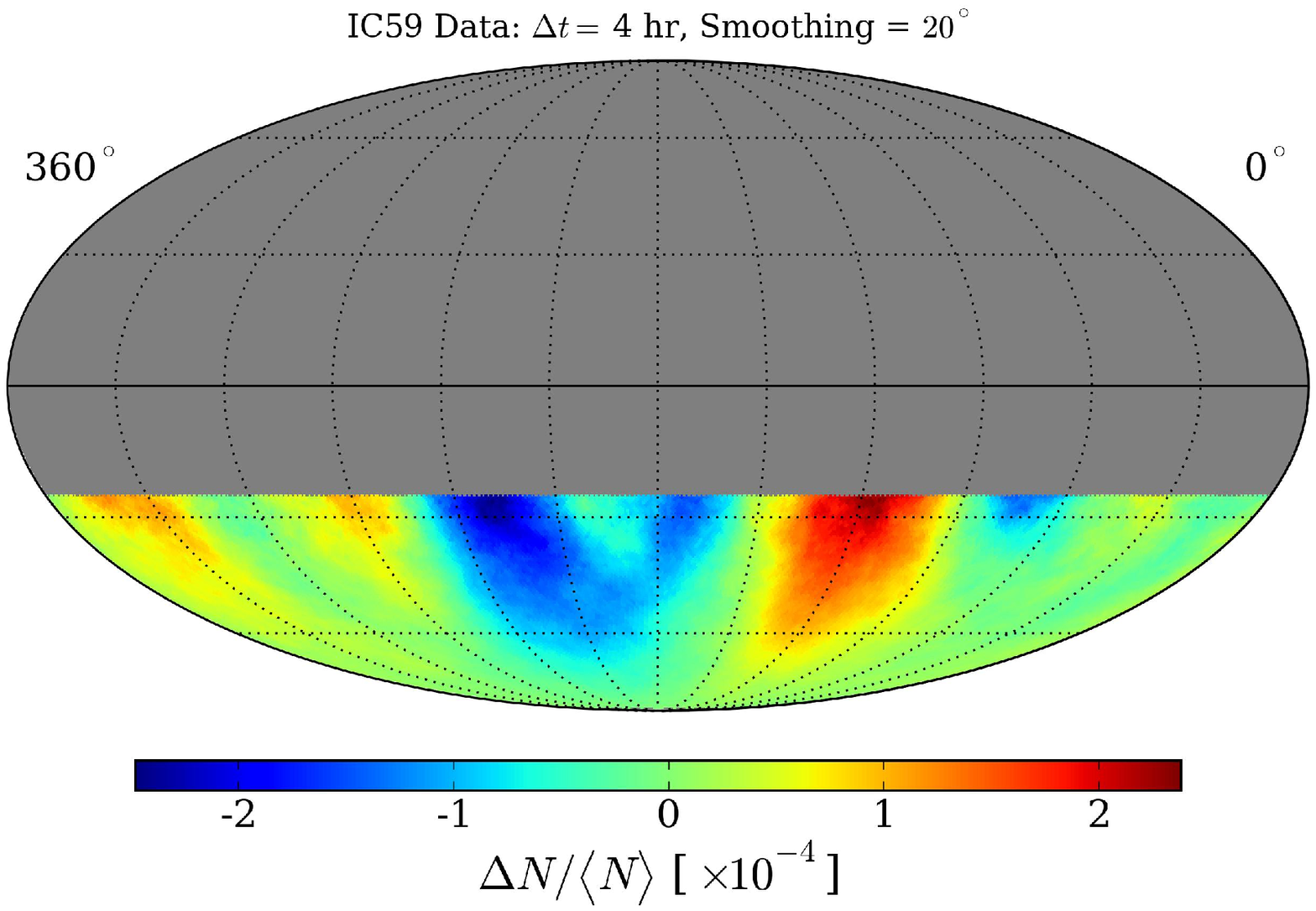}
   \includegraphics[width=.45\textwidth]{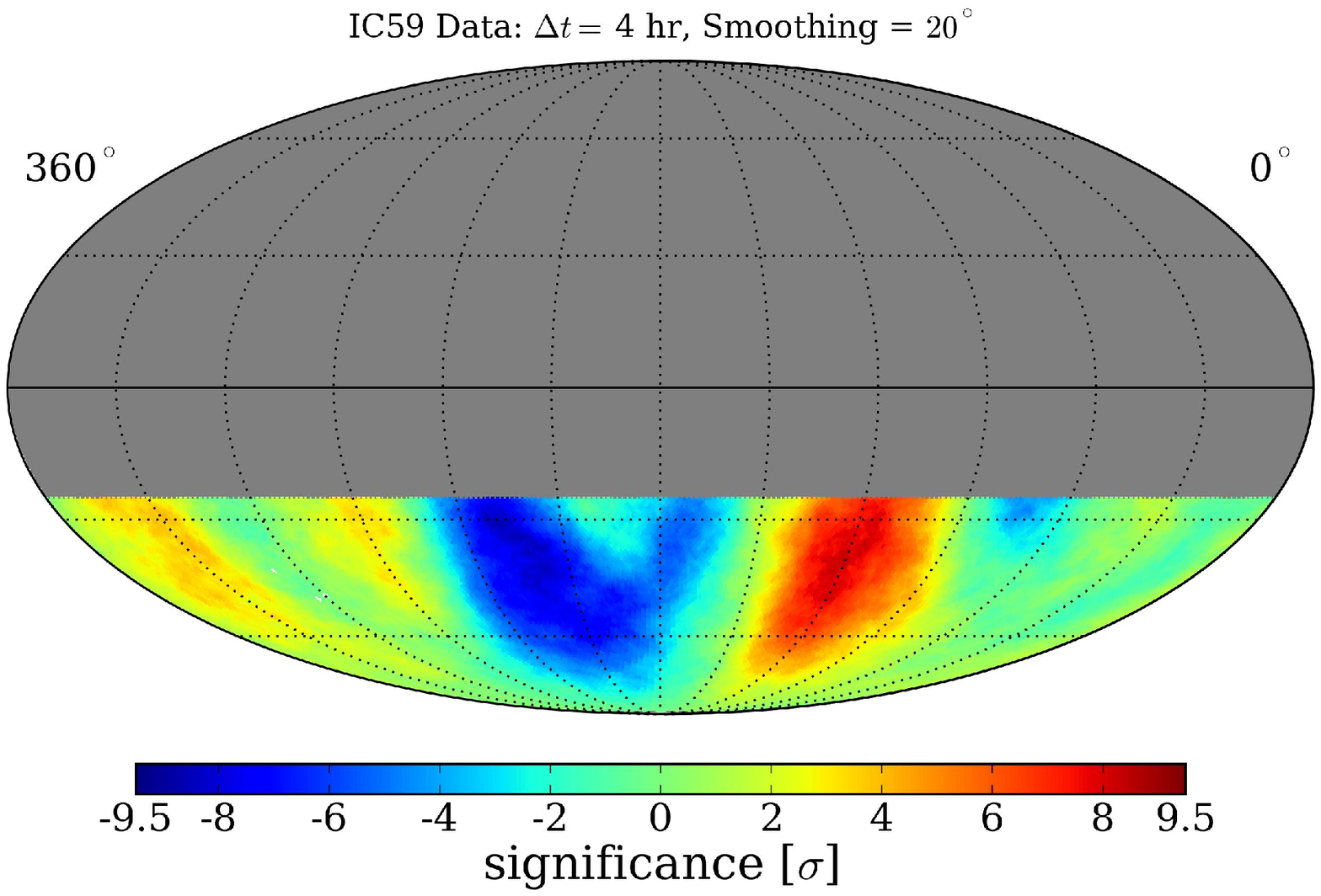}
\end{array}$
\end{center}
  \caption{Relative intensity {\em (left)} and significance {\em (right)} map in
    equatorial coordinates for $\Delta t=4$\,hours and an integration radius of 
    $20^\circ$.}
    \label{Fig:SignificanceMapSmoothed}
\end{figure}

Skymaps of the relative intensity and significance for $\Delta t=4$\,hours are
shown in Fig.\,\ref{Fig:SignificanceMapSmoothed}.  The maps have been smoothed
by $20^\circ$ to allow for a direct comparison with
Fig.\,\ref{fig:ic59FitdqResSmoothed}.  The most prominent features of the map
are a single broad excess and deficit, with several small excess regions
observed near the edge of the exposure region.  The broad excess is centered at
$\alpha=(121.7^{+4.8}_{-7.1})^\circ$ and $\delta=(-44.2^{+12.1}_{-7.8})^\circ$,
at the same position as Region 1 in Table~\ref{table:5sigmaspots}.  The optimal
smoothing scale of the excess is $25^\circ$, with a pre-trials significance of
$9.6\sigma$.  A second significant excess is observed at
$\alpha=(341.7^{+1.4}_{-5.6})^\circ$ and $\delta=(-34.9^{+3.6}_{-6.8})^\circ$
with a peak significance of $5.8\sigma$ at a smoothing scale of $9^\circ$.
This feature does not appear to have a direct match in
Fig.~\ref{fig:ic59FitdqResSmoothed}, but is roughly aligned in right ascension
with the excess identified in Table~\ref{table:5sigmaspots} as Region 4.  We
also note that the second-largest excess in Table~\ref{table:5sigmaspots},
Region 2, is visible near $\alpha=263.0^\circ$ in
Fig.~\ref{Fig:SignificanceMapSmoothed}, but with a pre-trials peak significance
of $4.5\sigma$ after smoothing by $13^\circ$.

The differences in significance between Figs.~\ref{fig:ic59FitdqResSmoothed}
and \ref{Fig:SignificanceMapSmoothed} can be attributed to the fact that some
contributions from the low-$\ell$ moments are still present in this
analysis.  The broad excess observed here is co-located with the maximum of the
large-scale structure shown in Fig.~\ref{fig:ic59_fitdq}, enhancing its
significance.  By comparison, the excess in Region 2 is close to the minimum of
the large-scale structure, weakening its significance.  The leakage of
large-scale structure into the $\Delta t=4$\,hour skymap also explains the
large deficit near $\alpha=220^\circ$; due to its co-location with the minimum
of the dipole and quadrupole, the size of the deficit is enhanced considerably.

\begin{figure}[t]
\begin{center}	
  \includegraphics[width=0.6\textwidth]{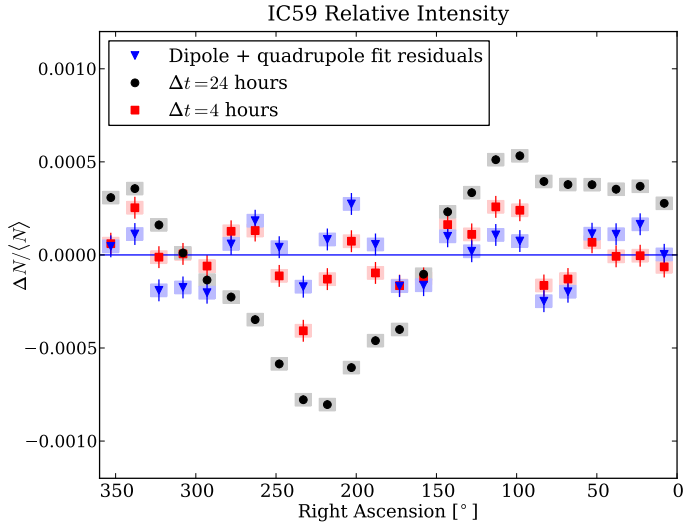}
\end{center}
  \caption{Relative intensity in the declination band
  $-45^\circ<\delta<-30^\circ$. The blue points show the result after 
  subtracting the dipole and quadrupole moments. The black points correspond 
  to \dt=24\,hours and show the large-scale structure, the red points
  correspond to \dt~=~4\,hours.    The error boxes
  represent systematic uncertainties.}
  \label{fig:ic59_relInt_stat_sys}
\end{figure}

This effect is illustrated in Fig.\,\ref{fig:ic59_relInt_stat_sys}, which shows
the relative intensity for the declination range $-45^\circ<\delta<-30^\circ$, 
projected onto the right ascension axis.  This declination range is chosen because
it contains some of the most significant structures of the skymaps.  The blue 
points show the relative intensity corresponding to Fig.\,\ref{fig:ic59FitdqResSmoothed}, 
{\it i.e.}, the skymap after subtraction of dipole and quadrupole moments.  The black 
and red points show the relative intensity for skymaps obtained with the method
described in this section; the black points correspond to $\Delta t=24$\,hours, 
the red points to $\Delta t=4$\,hours.  In the case of $\Delta t=24$\,hours, 
the large scale structure dominates.  For $\Delta t=4$\,hours, the large scale 
structure is suppressed, and the smaller features become visible.  The blue and
red curves show excesses and deficits at the same locations, but with different
strengths.  As the red curve still contains some remaining large scale structure,
maxima and minima are enhanced or weakened depending on where they are located
with respect to the maximum and minimum of the large-scale structure.
The systematic error for the relative intensity values in 
Fig.\,\ref{fig:ic59_relInt_stat_sys} is taken from the analysis of the data in 
anti-sidereal time as described in the next section.

\begin{figure}[t]
\begin{center}	
$\begin{array}{cc}	
  \includegraphics[width=.40\textwidth]{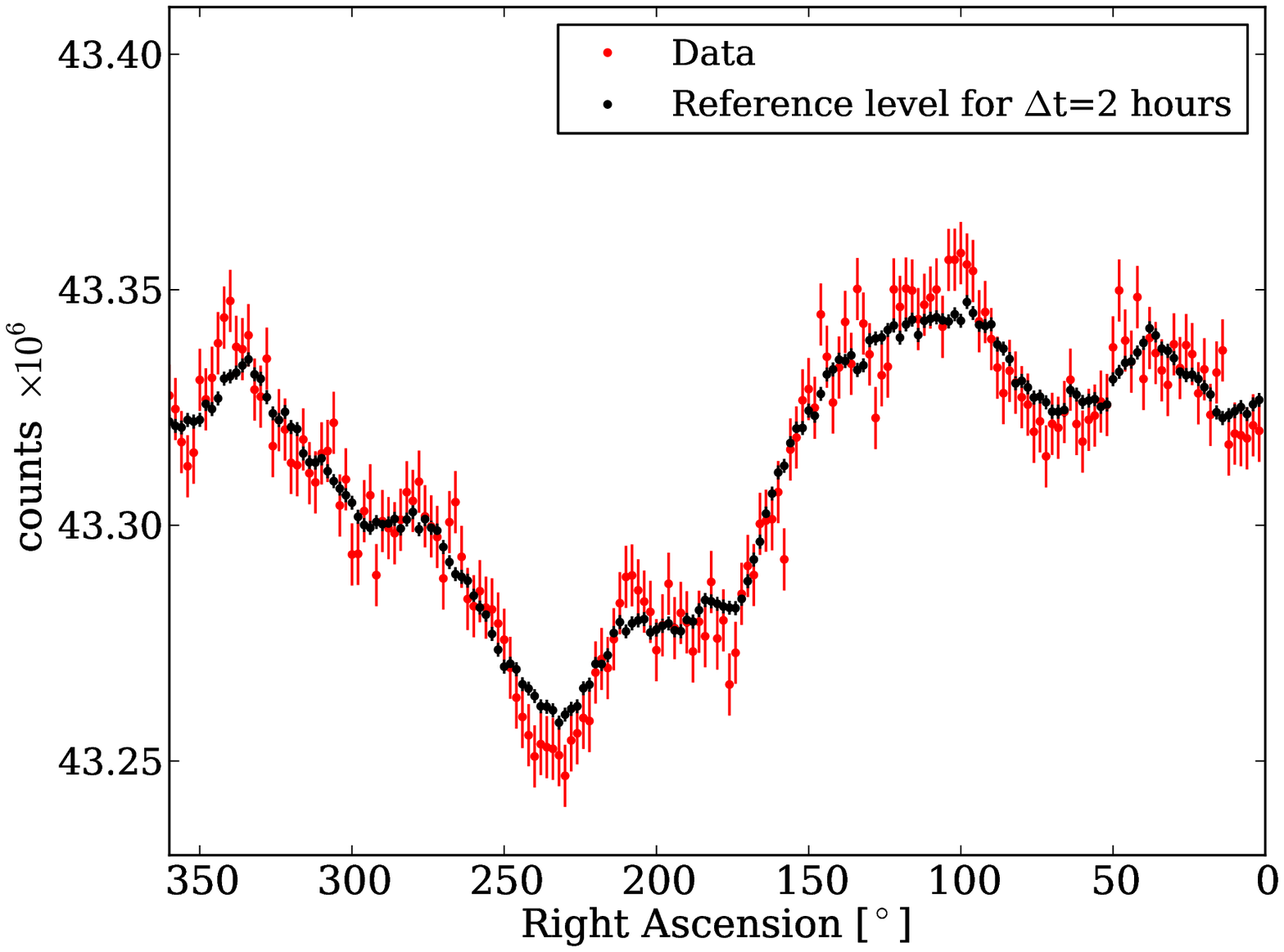}
  \includegraphics[width=.40\textwidth]{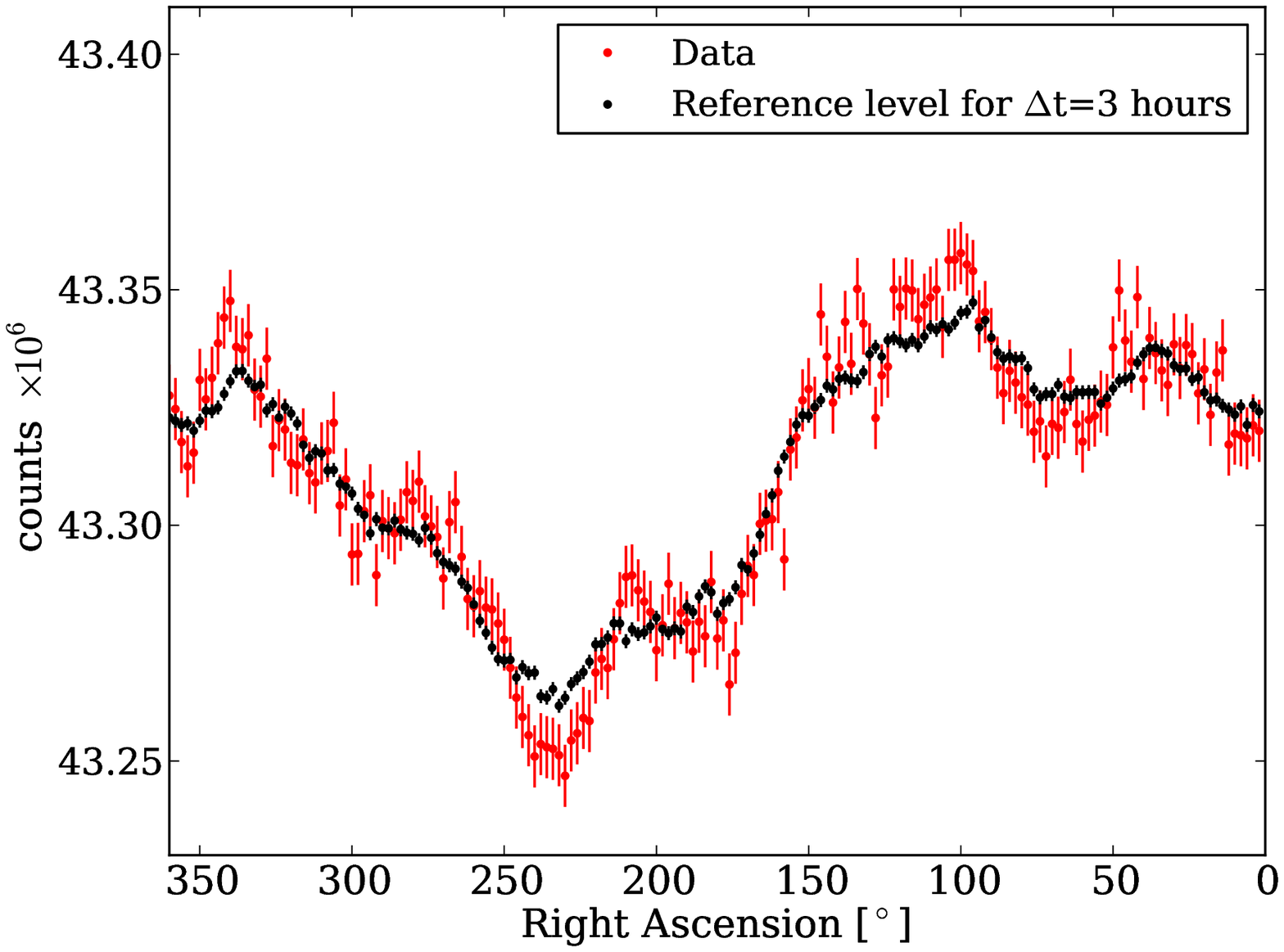}\\
  \includegraphics[width=.40\textwidth]{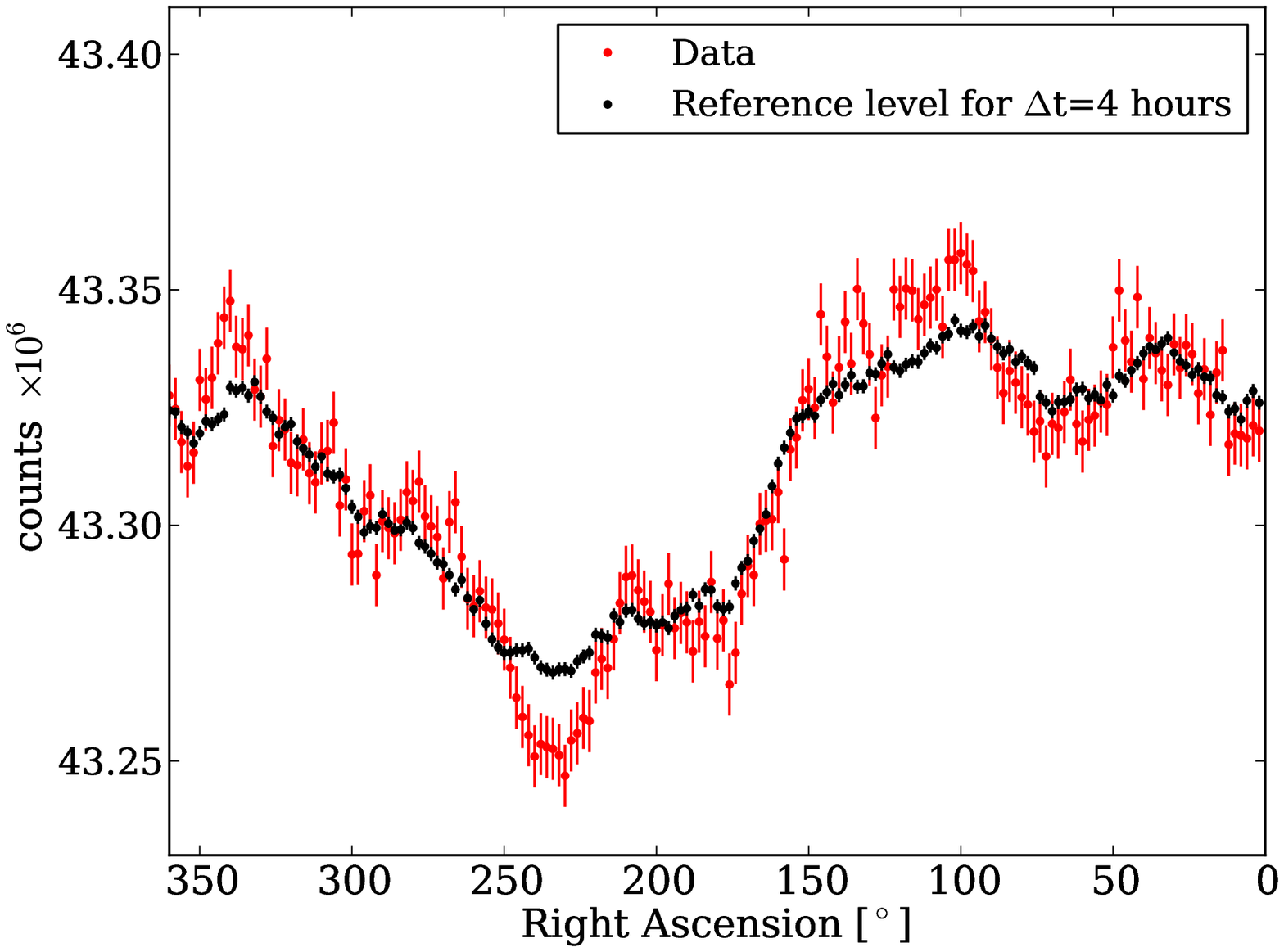}
  \includegraphics[width=.40\textwidth]{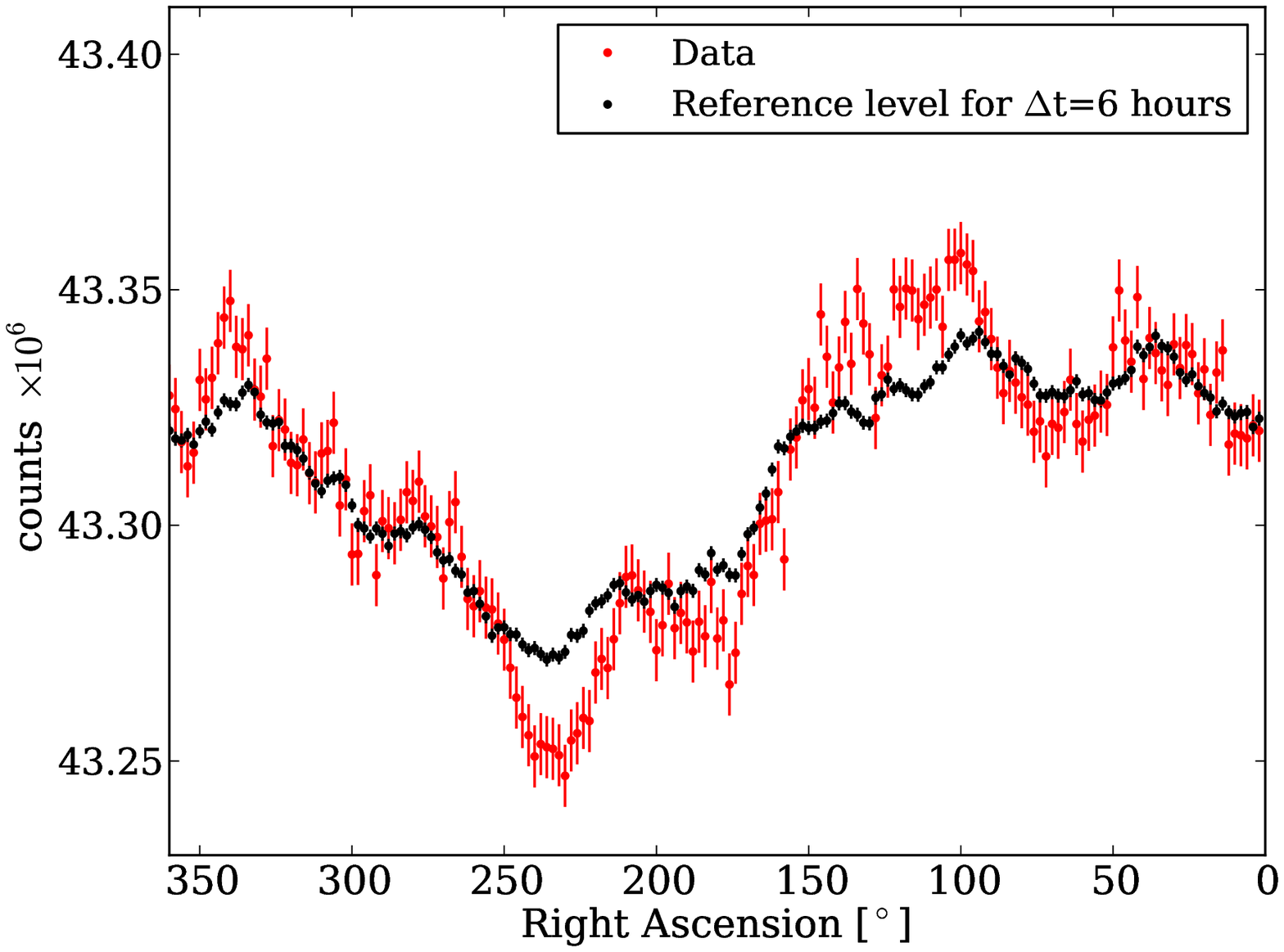}\\
  \includegraphics[width=.40\textwidth]{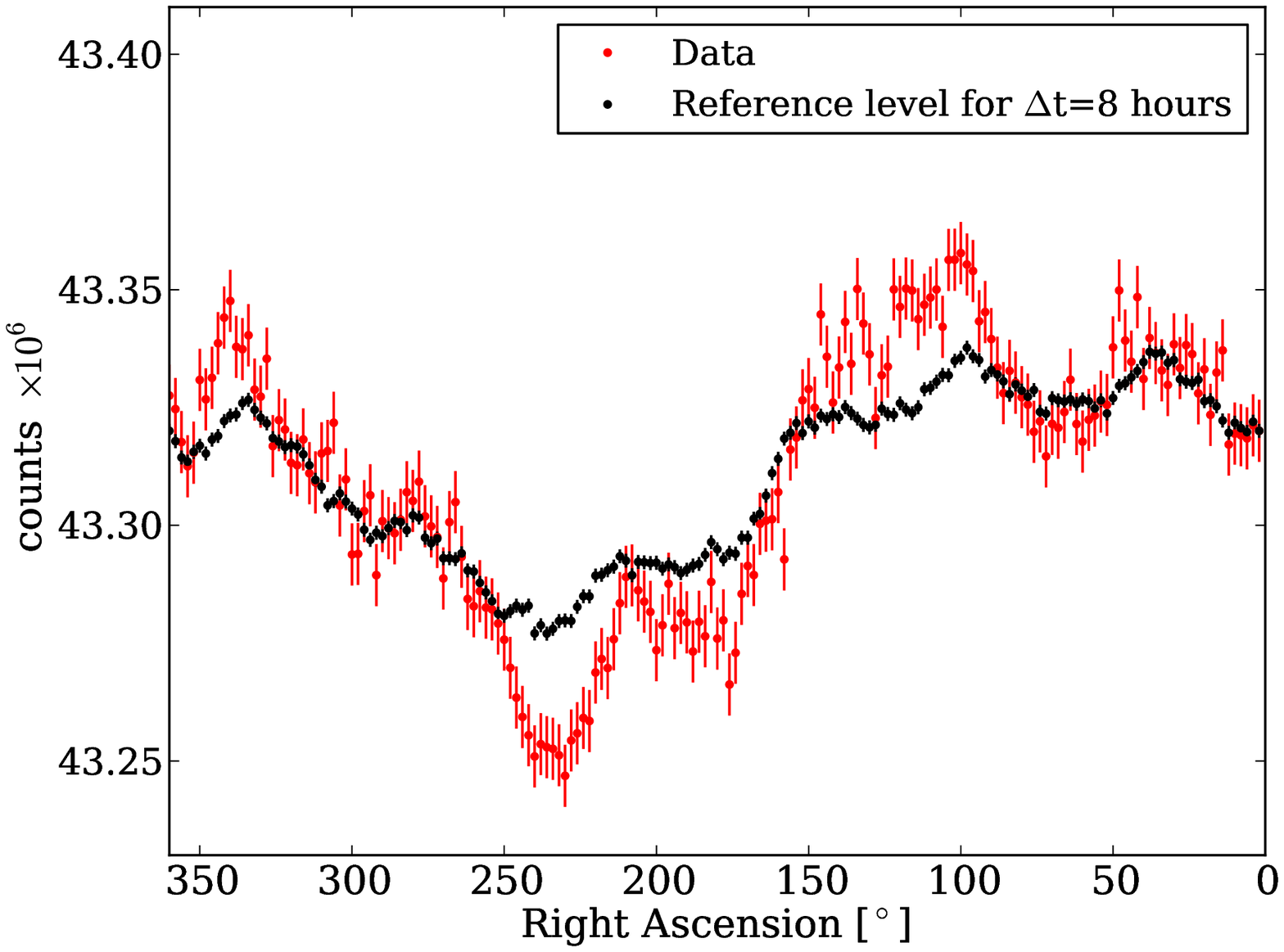}
  \includegraphics[width=.40\textwidth]{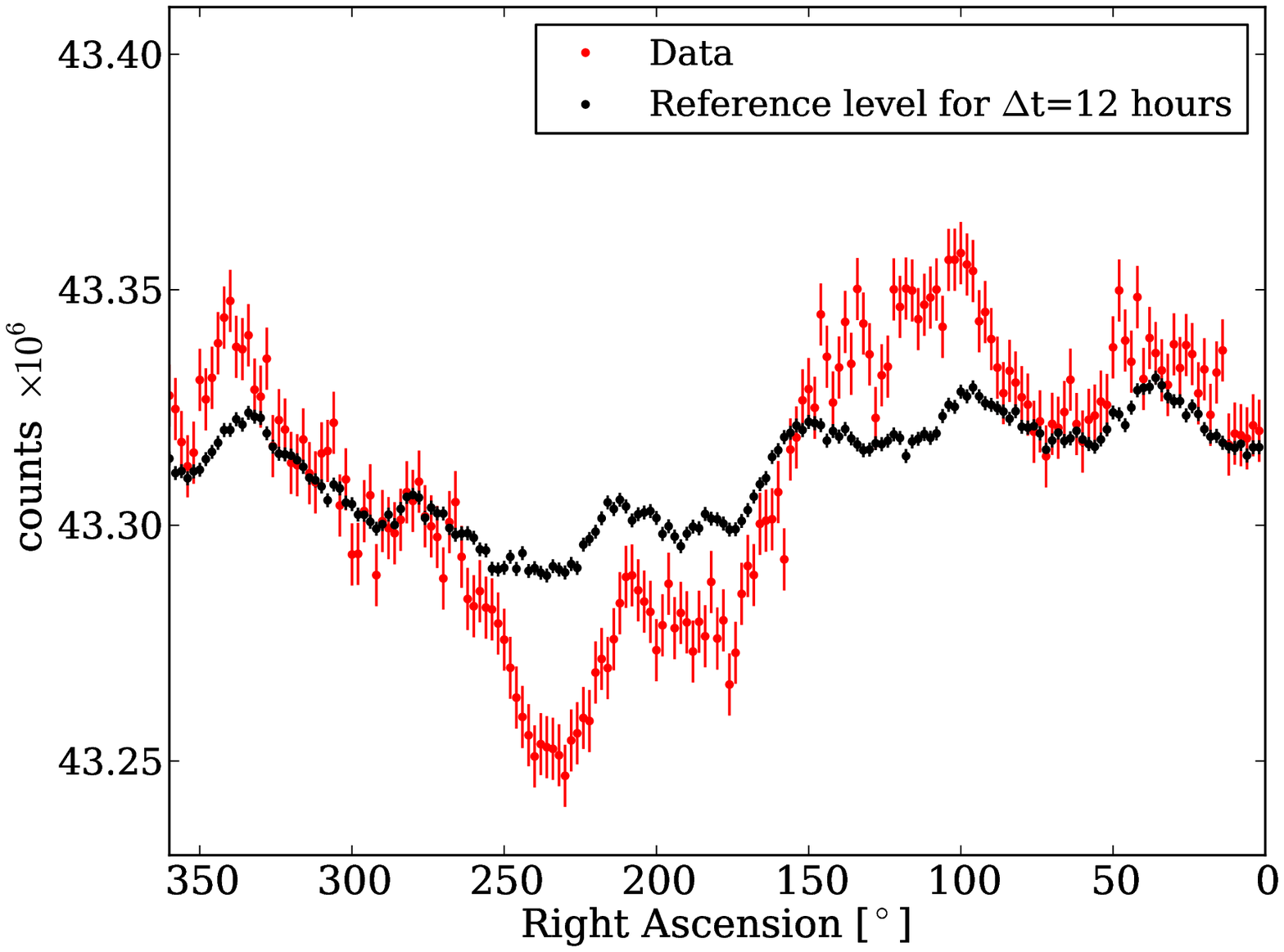}\\
  \includegraphics[width=.40\textwidth]{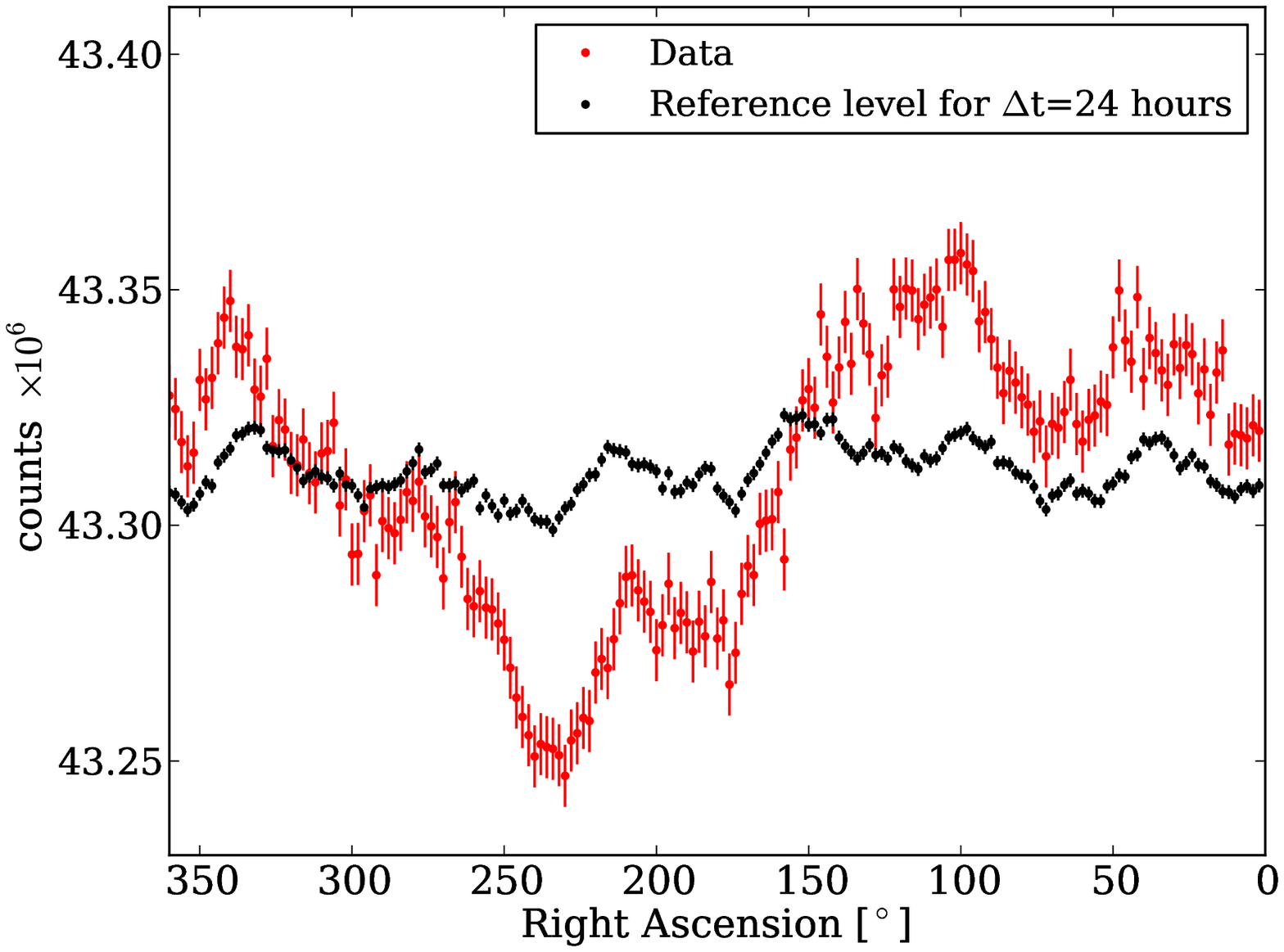}
\end{array}$
\end{center}
  \caption{Number of events (red) and reference level (black), 
  with statistical uncertainties, as a function of right ascension 
  for the declination range $-45^\circ<\delta<-30^\circ$. 
  The reference level is estimated in different time windows, from 2 hours ({\em top-left}) 
  to 24 hours ({\em bottom}). Each plot has been created using independent 
  $15^\circ\delta \times 2^\circ$ bins in right ascension.}
\label{Fig:SignalBGRAproj}
\end{figure}

Finally, we note that the presence of the small-scale structure can be 
verified by inspection of the raw event counts in the data.  
Figure~\ref{Fig:SignalBGRAproj} shows the observed and expected event 
counts for declinations $-45^\circ<\delta<-30^\circ$, projected onto 
the right ascension axis.  The seven panels of the figure contain the 
projected counts for seven time scrambling windows 
$\Delta t=\left\{2, 3, 4, 6, 8, 12, 24~\text{hours}\right\}$.  For small 
values of $\Delta t$, the expected counts agree with the data; for example, 
when $\Delta t=2$\,hours, the data exhibit no visible deviation from the 
expected counts.  For larger values of $\Delta t$, the expected count 
distribution flattens out as the technique to estimate the reference level
no longer over-fits the large structures.  When $\Delta t=24$\,hours, the 
reference level is nearly flat, and the shape of the large-scale 
anisotropy is clearly visible from the raw data.  

\clearpage

\section{Systematic Checks}\label{sec:systematics}

Several tests have been performed on the data to ensure the stability of the
observed anisotropy and to rule out possible sources of systematic bias.  Among
the influences that might cause spurious anisotropy are the
detector geometry, the detector livetime, nonuniform exposure of the
detector to different regions of the sky, and diurnal and seasonal variations
in atmospheric conditions.  Due to the unique location of the IceCube detector
at the South Pole, many of these effects play a lesser role for IceCube than
for detectors located in the middle latitudes.  The southern celestial sky is
fully visible to IceCube at any time, and changes in the event rate tend to
affect the entire visible sky.  Seasonal variations are of order
$\pm\,10\%$~\citep{Tilav:2010hj}, but the changes are slow and the reference level
estimation technique is designed to take these changes into account.  This is
also true for any effects caused by the asymmetric detector response due to the
geometrical configuration of the detector.  In this section, we test the accuracy
of these assumptions.

\subsection{Solar Dipole Analysis}\label{subsec:solarDipole}

As mentioned in Section~\ref{subsec:prevObs}, any observer moving through a
plasma of isotropic cosmic rays should observe a difference in intensity
between the direction of the velocity vector and the opposite direction.
Therefore, cosmic rays received on Earth should exhibit a dipole modulation in
solar time caused by the Earth's orbital velocity around the Sun.  The expected
change in the relative intensity is given by
\begin{equation}\label{eq:relIntCG}
  \frac{\Delta I}{\langle I\rangle}=
  \left(\gamma+2\right)\frac{v}{c}\cos\rho~~,
\end{equation}
where $I$ is the cosmic ray intensity, $\gamma=2.7$ the power law index of 
the cosmic ray energy spectrum, $v/c$ the ratio of the Earth's velocity with 
respect to the speed of light, and $\rho$ the angle between the cosmic ray
arrival direction and the direction of motion~\citep{Gleeson:1968ax}.  With a 
velocity of $v=30\,\mathrm{km}\,\mathrm{s}^{-1}$, the expected amplitude is 
$4.7\times10^{-4}$.  Note that the power law spectral index has a systematic 
uncertainty (see for example~\cite{Biermann:2010qn} for a discussion) and the 
Earth's velocity is not precisely constant, but both of these uncertainties 
are too small to be relevant in our comparison of the predicted dipole strength 
to the measured strength.  The solar dipole effect has been measured with several 
experiments~\citep{Amenomori:2004bf, Amenomori:2007ug,Abdo:2008aw} and 
provides an important check of the reliability of the analysis techniques 
presented earlier, as it verifies that the techniques are sensitive to a 
known dipole with an amplitude of roughly the same size as the 
structures in the equatorial skymap.

In principle, the solar dipole is not a cause of systematic uncertainties in 
the analysis of cosmic ray anisotropy in sidereal time (equatorial coordinates).  
The solar dipole is visible only when the arrival directions are plotted in a
frame where the Sun's position is fixed in the sky.  A signal in this
coordinate system averages to zero in sidereal time over the course of one
year.  However, any seasonal variation of the solar dipole can cause a spurious 
anisotropy in equatorial coordinates.  The effect works both ways: a seasonal
variation in the sidereal anisotropy will affect the solar dipole.  A standard
way to study the extent of these contaminations is by use of two artificial time 
scales, anti-sidereal and extended-sidereal time.  Anti-sidereal time is calculated
by reversing the sign of the transformation between universal time and sidereal 
time. 
Each sidereal day is slightly shorter than the solar day (universal time) by about 4 minutes, 
while each anti-sidereal day is longer than a solar day by the same factor.
Anti-sidereal time therefore has 364.25 days ({\it i.e.}\ complete revolutions in the coordinate frame)
per year, one day less than the solar year (365.25 days) and two days less than the 
sidereal year (366.25 days). 
Similarly, each extended sidereal day is shorter than a sidereal day by about 4 minutes 
(8 minutes shorter than the solar day).
Extended sidereal time has therefore 367.25 days per year.  No physical phenomena 
are expected to occur in the anti-sidereal or in the extended-sidereal 
frame.  However, systematic distortions in the sidereal anisotropy due to seasonal
variations of the solar dipole will produce a ``signal'' in anti-sidereal time.
Similarly, distortions in the solar dipole due to seasonal variations of the 
sidereal anisotropy will produce a ``signal'' in extended-sidereal time.  We follow
the example of~\cite{Amenomori:2007ug} and~\cite{Abdo:2008aw} and use anti-sidereal
time for an estimate of the error from seasonal variations on the amplitude of the 
sidereal anisotropy, and extended-sidereal time to estimate the systematic error
on the solar dipole amplitude.

\begin{figure}[t]
  \begin{center}
    \includegraphics[width=0.7\textwidth]{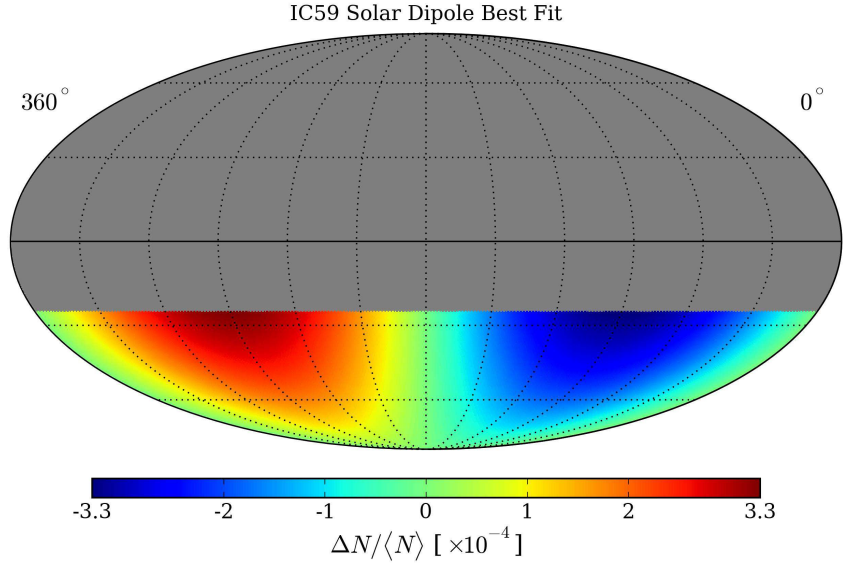}
    \caption{Best fit results to the IC59 data expressed in solar coordinates.
    In this coordinate system, the velocity vector of the motion of the Earth
    about the Sun is oriented at a longitude of $270^\circ$.}
    \label{fig:solarDipoleFit}
  \end{center}
\end{figure}

To measure the solar dipole anisotropy we estimate the reference level using
a time window $\Delta t=$\,24 hours, which maximizes the sensitivity to
large-scale features.  The data and reference maps are produced in a
coordinate system where the latitude coordinate is declination and the
longitude coordinate represents the angular distance from the Sun in right
ascension, defined as the difference between the right ascension of each event
and the right ascension of the Sun.  In this coordinate system the Sun's
longitude is fixed at $0^\circ$ and we expect, over a full year, an excess in
the direction of motion of the Earth's velocity vector (at $270^\circ$) and a
minimum in the opposite direction. 

\begin{table}[bt]
  \begin{center}
  {\small
     \begin{tabular}{cr} \hline
      Coefficient  & \multicolumn{1}{c}{Value (stat. + syst.)}\\
            $\quad$      & \multicolumn{1}{c}{($\times10^{-4}$)}\\
       \hline
      $m_0$ &  $-0.03\pm0.06\pm0.02$ \\
      $p_x$ &  $ 0.02\pm0.14\pm0.97$ \\
      $p_y$ &  $-3.66\pm0.14\pm0.17$ \\
      $p_z$ &  $-0.03\pm0.07\pm0.01$ \\
      \hline
    \end{tabular}
  }
  \caption{\label{table:solarDipoleFit}
    Coefficients of a dipole and constant offset fit to the IC59 solar
    coordinate data.  The systematic error on the fit parameters is 
    estimated using the results of a fit using extended-sidereal time as described in the text.}
  \end{center}
\end{table}

The data are fit using the dipole and quadrupole expansion given in
Eq.~\eqref{eq:dqfit}.  The quadrupole coefficients are found to be equivalent
to zero within the fit uncertainties, so the fit is repeated with only a dipole
term and a constant offset.  The dipole describes the data well; the fit
$\chi^2/\text{ndf}=14207 / 14192$ corresponds to a $\chi^2$-probability of
$41.6\%$.  The best fit coefficients are listed in
Table~\ref{table:solarDipoleFit}.  Only one free parameter, the $p_y$ component
of the dipole fit, differs significantly from zero.
Hence, the dipole is aligned at a longitude of $270^\circ$ within the
equatorial plane of this coordinate system, following the expectation for a
dipole in the cosmic ray skymap caused by relative motion about the Sun.

The amplitude of the dipole is
$(3.66\pm0.14_\text{stat}\pm0.99_\text{sys})\times10^{-4}$.
The systematic uncertainty is evaluated by fitting a dipole to the data in a 
coordinate system using extended-sidereal time.  We have conservatively estimated 
this systematic uncertainty by taking the amplitude of the
dipole in extended-sidereal coordinates.  Within the large systematic error,
the amplitude of the solar dipole agrees with the prediction.  A more detailed
study of the solar dipole anisotropy in IceCube data will follow in a separate
publication.

\subsection{Anti-Sidereal Time Analysis}\label{subsec:antiSidereal}

As described in the previous section, we use the analysis of the data in 
the anti-sidereal time frame to study systematic effects caused by seasonal 
variations.  For this test, we produce skymaps where anti-sidereal time
is used instead of sidereal time in the coordinate transformation 
from local detector coordinates to ``equatorial'' coordinates.  
Skymaps produced in this way are subjected to the same analyses as the
true equatorial maps.  Neither the angular power spectrum nor the
skymaps show any significant deviation from isotropy.  In particular,
no regions of significant excess or deficit are observed in the anti-sidereal 
skymaps for any smoothing scale.  The systematic error bars shown in 
Fig.\,\ref{fig:ic59_relInt_stat_sys} are estimated by using the variation 
in anti-sidereal time as a measure of this error.

\subsection{Comparison with IC22 and IC40}\label{subsec:ic22ic40comp}

An important cross-check of the structure seen in the IC59 data set can be
made by applying the IC59 analysis to data recorded in the two data periods
prior to IC59.  The IC22 data set contains 5 billion events recorded between
July 2007 and April 2008, and the IC40 data set contains 19 billion events
recorded between April 2008 and May 2009.  While these data sets are smaller
than the IC59 data set due to the smaller detector size, we nevertheless expect 
to observe the most prominent structures in these data, albeit with reduced 
significance.

The IC22 and IC40 data can be used to verify that the structures observed in
the arrival direction distribution do not depend on the geometry of the
detector or the data taking period.  The shapes of both detector configurations 
are highly asymmetric, with a long axis and a short axis.  The asymmetry 
introduces a trigger bias into the data, because muon tracks aligned with the 
long axis are much more likely to satisfy the simple majority trigger conditions 
than events arriving along the short axis.  As a result, the local arrival direction
distribution of the IC22 and IC40 data is highly nonuniform in azimuth.

\begin{figure}[t]
\begin{center}
$\begin{array}{cc}  
  \includegraphics[width=.495\textwidth]{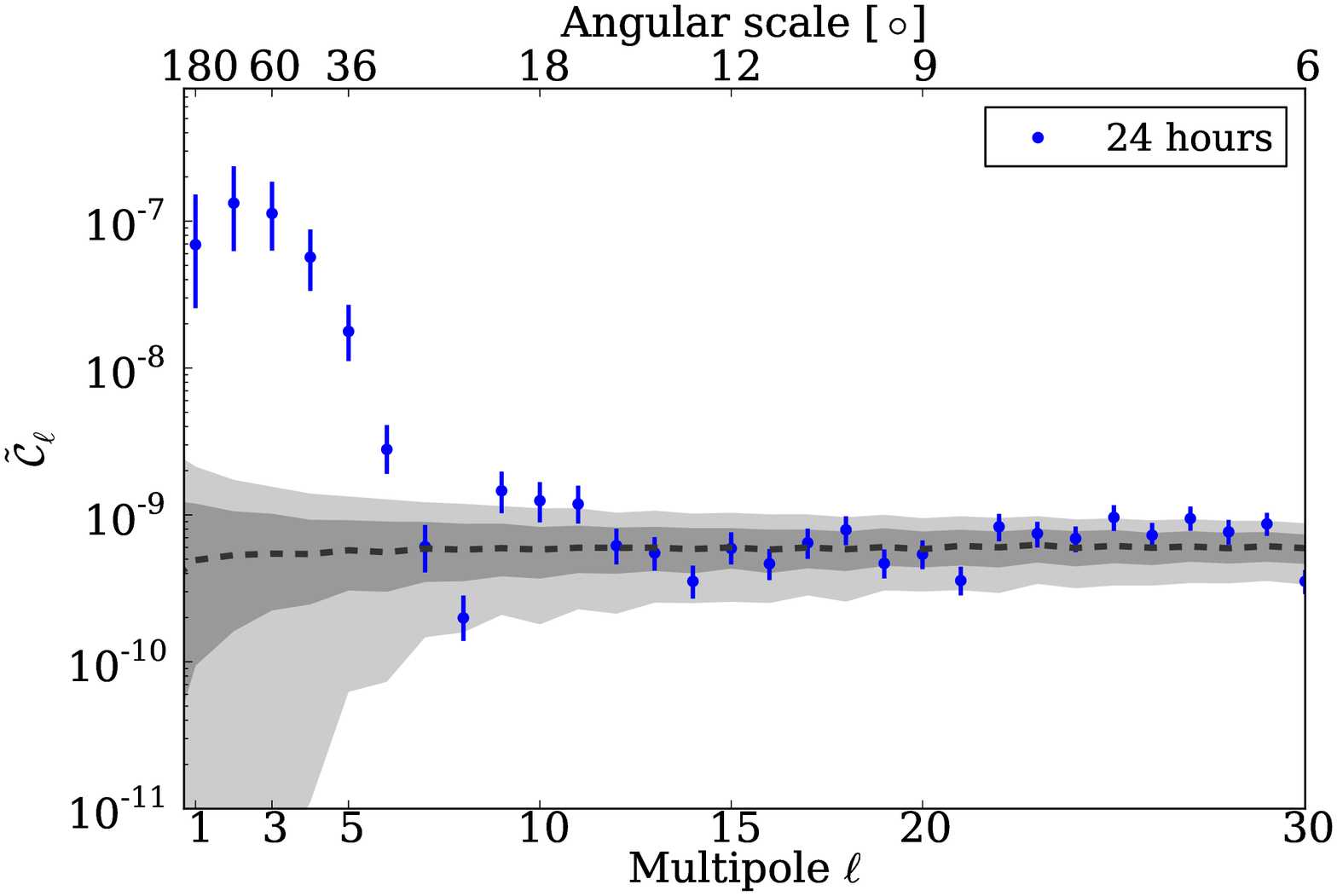}
  \includegraphics[width=.495\textwidth]{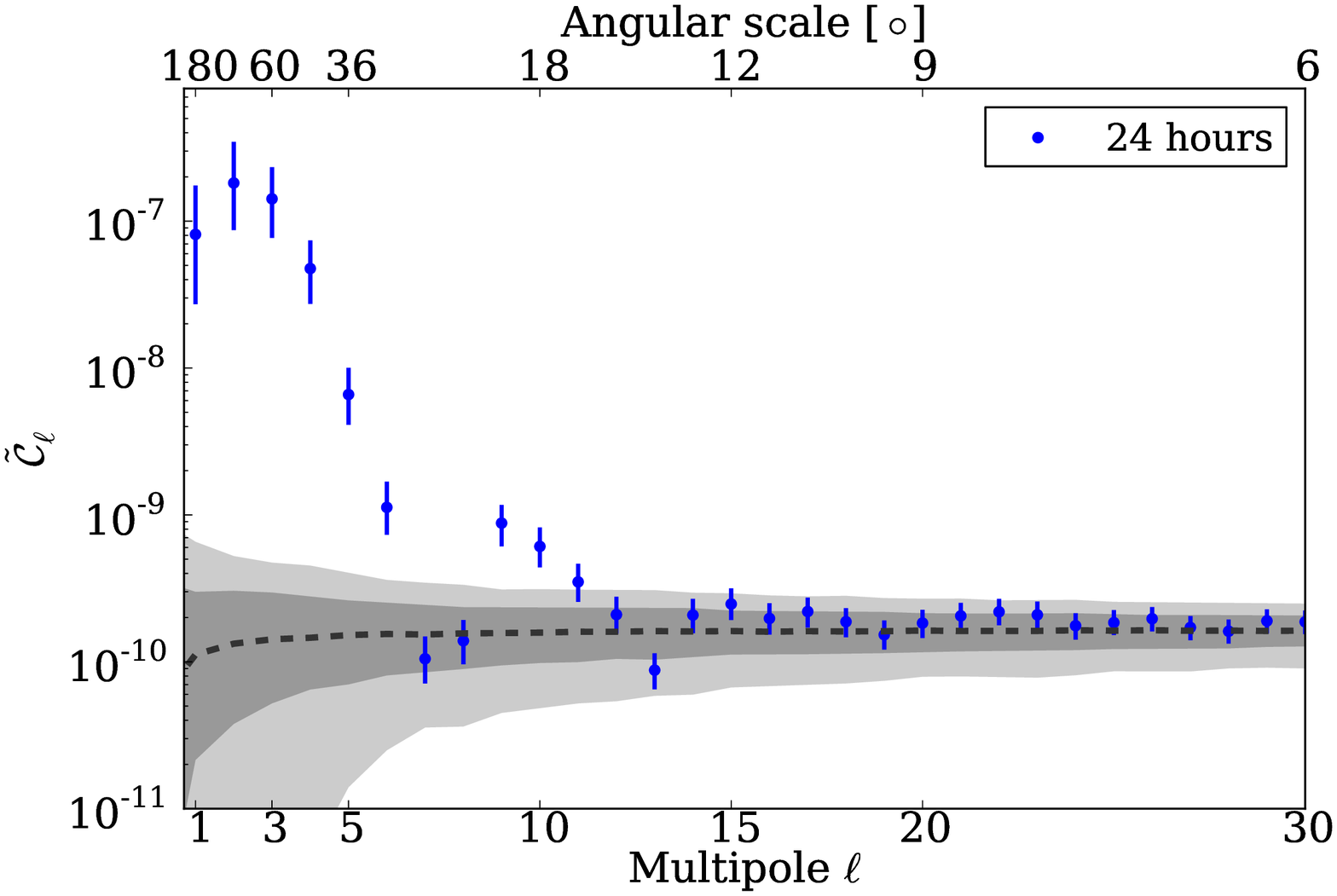}
\end{array}$
\end{center}
  \caption{ Angular power spectra for the relative intensity maps from IC22 (left),
and IC40 data (right). Errors bars are statistical. 
The gray bands indicate the distribution of the power spectra in a large 
    sample of isotropic data sets, showing the 68\% (dark) and 95\% (light)
spread in the ${\cal \tilde{C}}_{\ell}$. }
  \label{fig:ps_ic22ic40}
\end{figure}

We repeat the main analysis steps described in Sec.\,\ref{sec:Analysis}.
Fig.\,\ref{fig:ps_ic22ic40} shows the angular power spectrum for 
IC22, IC40, and IC59.  Both small- and large-scale structures are present 
in all three data sets.

\begin{figure}[t]
  \begin{center}	
    \includegraphics[width=0.495\textwidth]{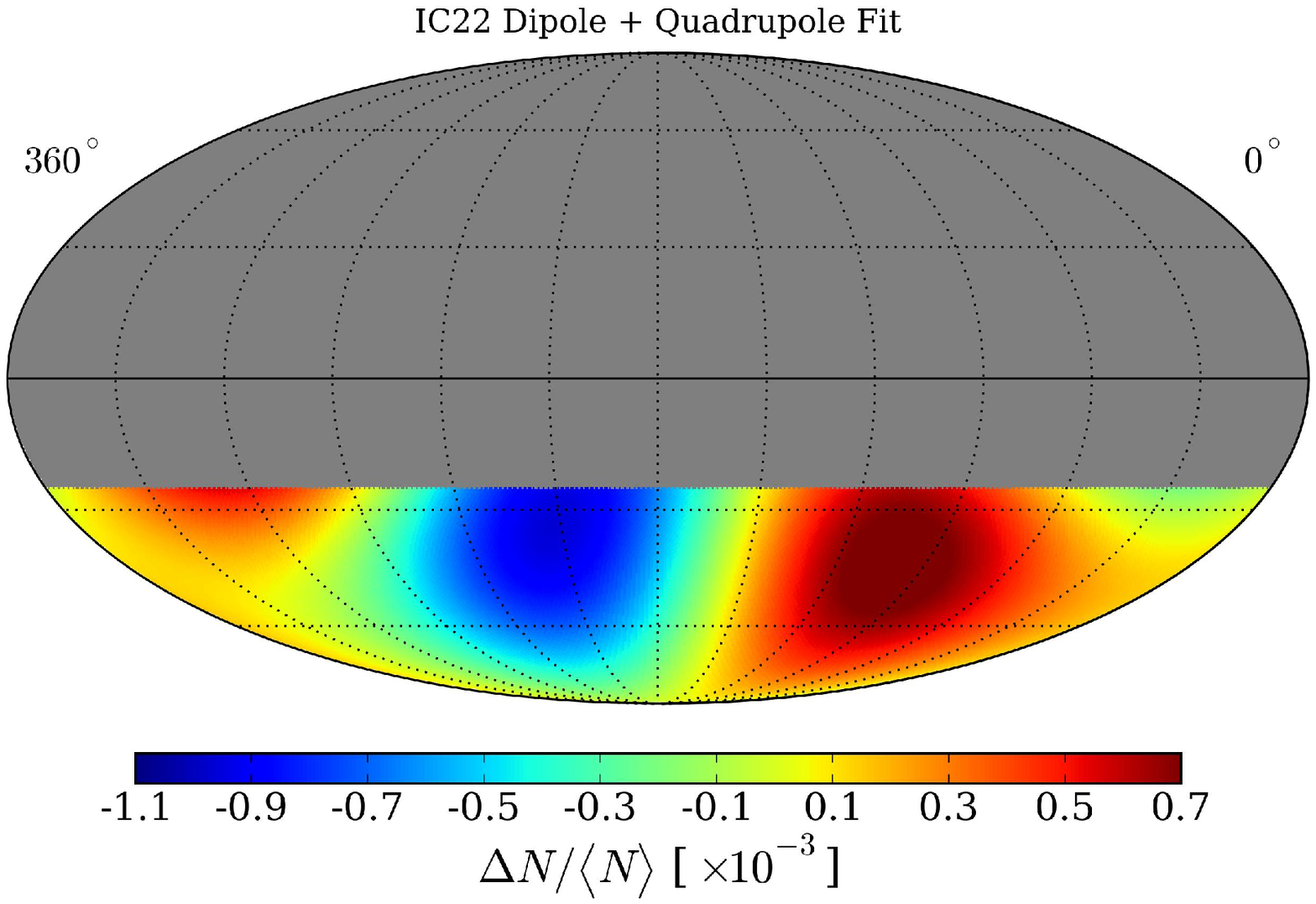}
    \includegraphics[width=0.495\textwidth]{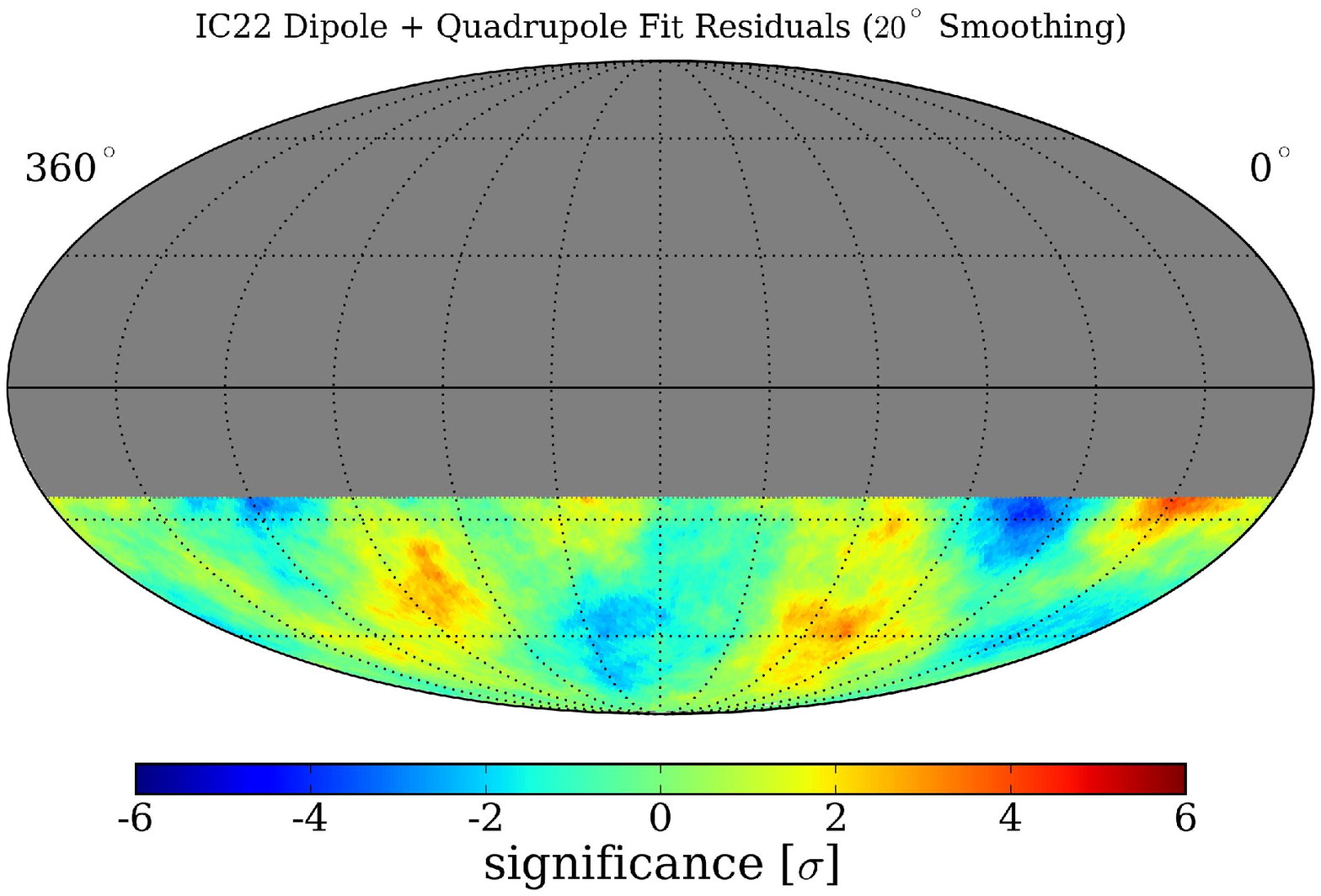} \\
    \includegraphics[width=0.495\textwidth]{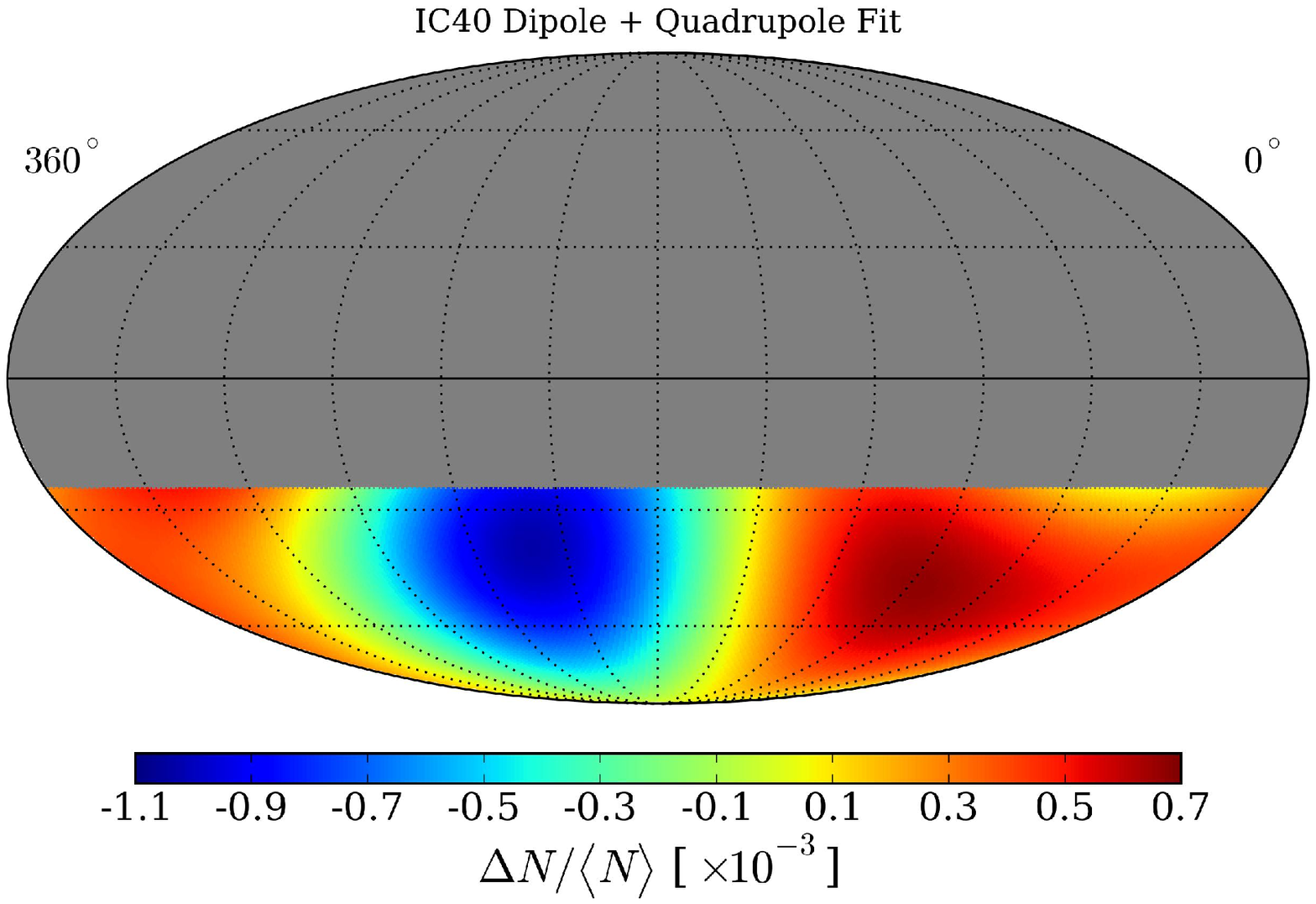}
    \includegraphics[width=0.495\textwidth]{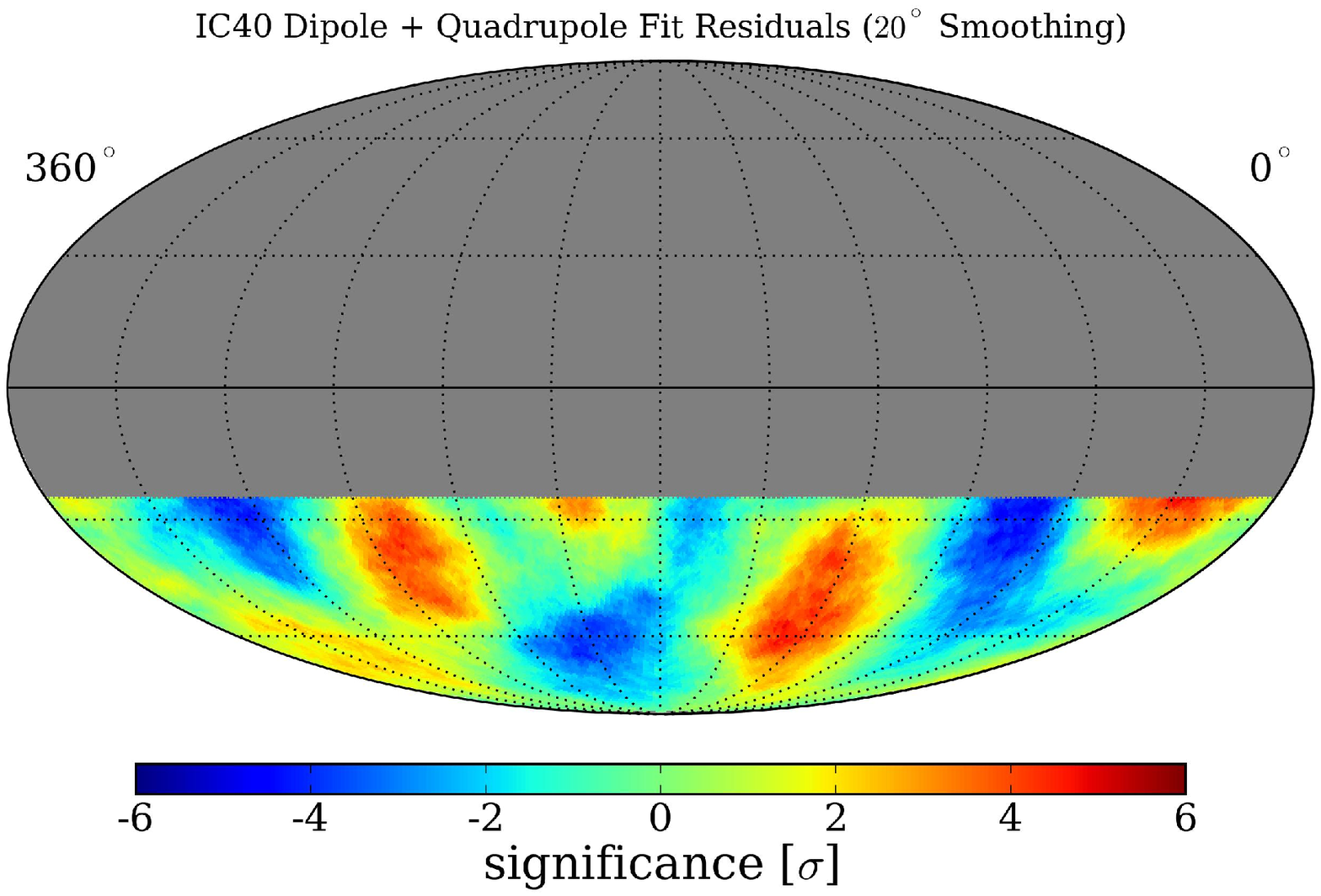}
    \caption{
      {\em Top}: Combined dipole and quadrupole fit of eq.~\eqref{eq:dqfit} to
      data from IC22 {\em (left)} and fit residuals after $20^\circ$
      smoothing {\em (right)}.
      {\em Bottom}: Dipole and quadrupole fit to data from IC40 {\em (left)} and fit
      residuals {\em (right)}.  
	\label{fig:IC22IC40}}
  \end{center}
\end{figure}

Fig.\,\ref{fig:IC22IC40} shows the result of the dipole and quadrupole 
fits (left) and the residual map after subtraction of dipole and quadrupole 
(right) for IC22 (top) and IC40 (bottom).  The residual maps are smoothed 
with a $20^\circ$ radius so they can be directly compared to 
Fig.\,\ref{fig:ic59FitdqResSmoothed}.  While none of the features in IC22
and IC40 have a pre-trials significance above $5\sigma$, they align with 
the regions of deficit and excess observed with IC59 data (cf. 
Fig.~\ref{fig:ic59FitdqResSmoothed}).  The main features on both small and 
large scales appear to be persistent in all data sets. 

\begin{figure}[t]
\begin{center}	
  \includegraphics[width=0.495\textwidth]{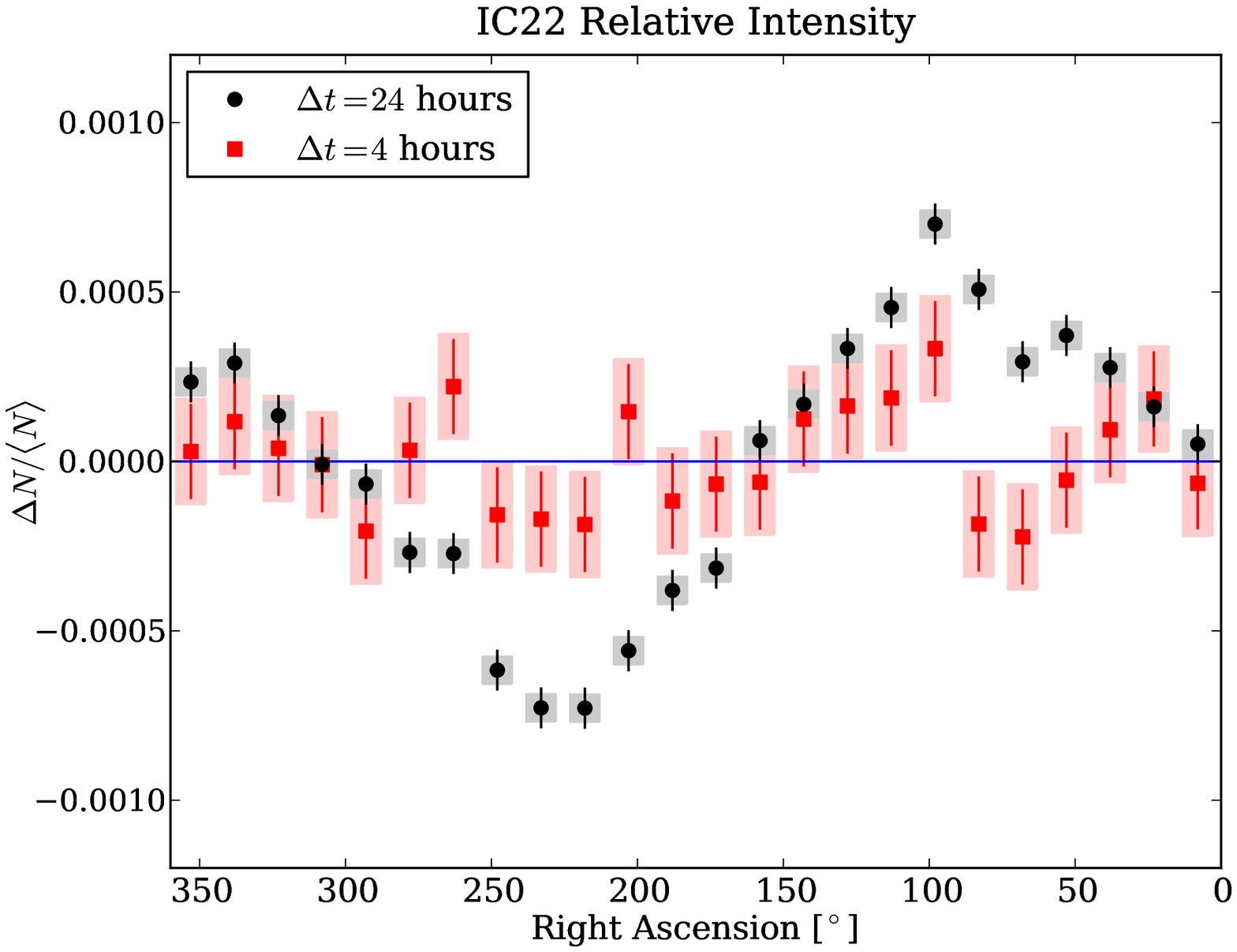}
  \includegraphics[width=0.495\textwidth]{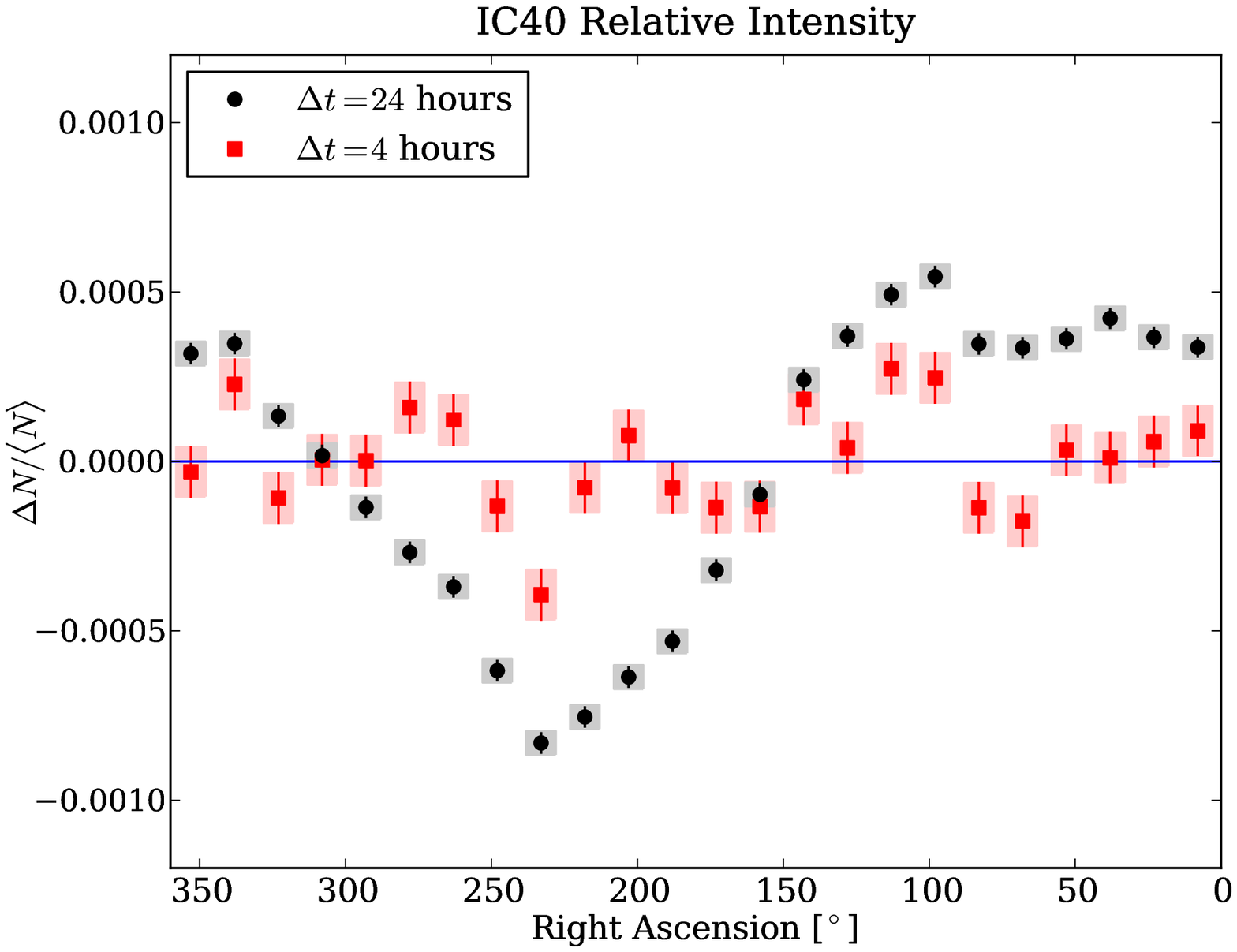}
\end{center}
	\caption{Relative intensity in the declination band
  between $-45^\circ$ and $-30^\circ$ for $\Delta t = 4$ hours for data from 
  IC22 {\em (left)} and IC40 {\em (right)}. 
  Statistical and systematic uncertainties are shown, with systematics
  calculated from the relative intensity distribution in anti-sidereal
  coordinates.
	\label{Fig:IC22IC40IC59}}
\end{figure}

Fig.\,\ref{Fig:IC22IC40IC59} compares the results of the analysis described in
Sec.\,\ref{subsec:timewindows} for the IC22 and IC40 data.  The figure shows
the relative intensity as a function of right ascension for the declination
band between $-45^\circ$ and $-30^\circ$, where the most significant deviations
from isotropy are found.  The systematic error band is estimated from the
relative intensity distribution in anti-sidereal time as described in
Sec.\,\ref{subsec:antiSidereal}.  The results for IC22 (left) and IC40 (right)
show that similar deviations are present in the IC22, IC40, and IC59 data,
again with increasing significance due to the increasing size of the data sets.

The stability of the results over several years of data taking and three
different detector configurations indicates that the anisotropy is not
produced by the geometry of the detector.  Since the temporal distribution of
detector livetime is also different for all three data sets, the stability of
the results indicates that the anisotropy is not affected by nonuniformities
in the detector livetime.  As expected, the time scrambling method accounts for
this effect.

\section{Conclusions}\label{sec:Conclusions}

Using 32 billion events recorded with the partially-deployed IceCube 
detector between May 2009 and May 2010, we have shown that the arrival 
direction distribution of cosmic rays with a median energy of 20\,TeV 
exhibits significant anisotropy on all scales up to $\ell=12$ in the
angular power spectrum.  The power spectrum is dominated by a dipole
and quadrupole moment, but also indicates the presence of significant 
structure on angular scales down to about $15^\circ$.  These
structures become visible in the skymap when the dominant dipole and
quadrupole moments are either subtracted or suppressed.  The residual
skymap shows both significant excesses and deficits, with the most
important excess reaching a post-trial significance of $5.3\,\sigma$ in IC59.
The relative intensity of the smaller-scale structures are about a 
factor of 5 weaker than the dipole and quadrupole structure.
A study of data taken with the smaller IC22 and IC40 detectors in
previous years confirms that these deviations from an
isotropic flux are consistently present in all data sets.

\begin{figure}[t]
  \begin{center}	
    \includegraphics[width=0.9\textwidth]{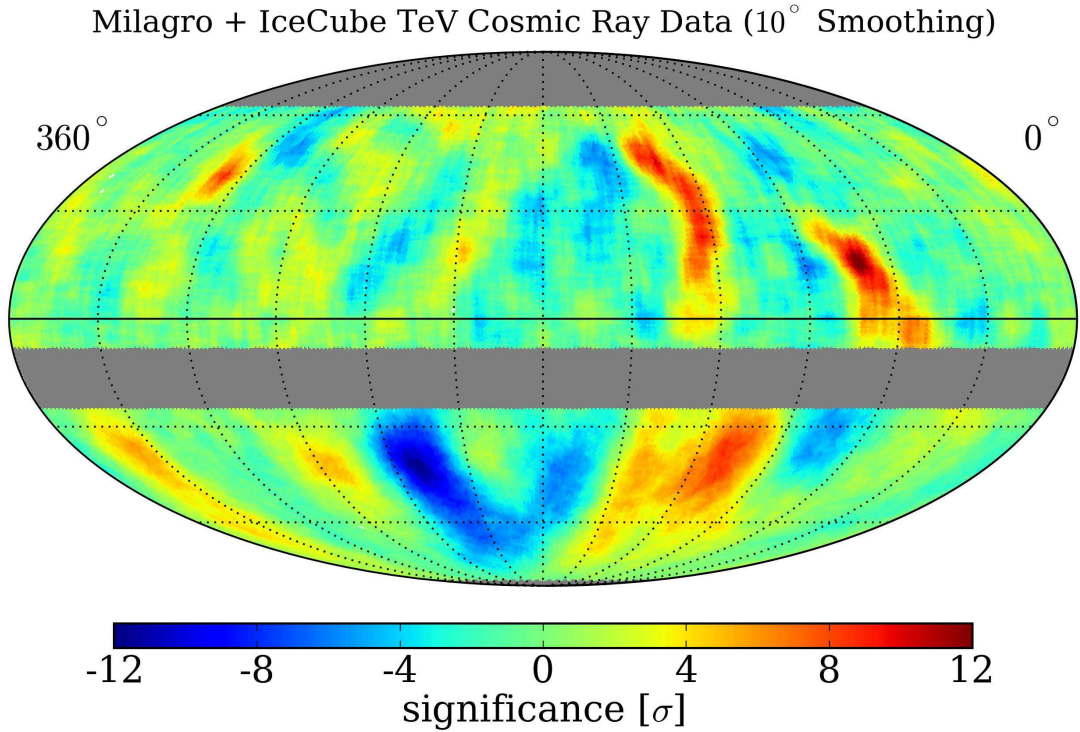}
    \caption{Combined map of significances in the cosmic ray arrival
    direction distribution observed by Milagro in the northern    
    hemisphere~\citep{Abdo:2008kr} and IceCube in the southern hemisphere
    (this analysis).  Both maps have been smoothed with a $10^\circ$ radius.
    \label{fig:ic_plus_milagro}}
  \end{center}
\end{figure}

Together with data from the $\gamma$-ray experiments in the northern
hemisphere, we now have an almost complete cosmic ray map of the entire sky at
TeV energies.  Fig.\,\ref{fig:ic_plus_milagro} shows the combined IceCube and
Milagro skymaps of small-scale anisotropy.  For this map, all available IceCube
data (IC22, IC40, and IC59) have been used, with a total of $5.6\times10^{10}$ events, 
and the analysis is performed using
the method described in Sec.\,\ref{subsec:timewindows} with a smoothing radius
of $10^\circ$ to match the Milagro analysis.  The combined skymap shows
significant excess regions in both hemispheres.  It is possible that the
structure around right ascension $120^\circ$ spans both hemispheres, as the
drop in significances around declination $\delta=0^\circ$ could be an artifact
of the smaller exposure of both detectors near $\delta=0^\circ$, which
corresponds to a region close to the horizon for both detectors.

There is currently no explanation for these local enhancements in the cosmic
ray flux.  We note that the two most significant excess regions in the 
southern sky (regions 1 and 2 in Tab.\,\ref{table:5sigmaspots}) are both 
located near the Galactic plane.  In addition, the position of one of the 
excess regions (region 1) coincides with the location of the Vela
pulsar at $(\alpha=128.8^\circ,\delta=-45.2^\circ)$.  At a distance of about 
300\,pc~\citep{Caraveo:2001ud},
Vela is one of the closest known supernova remnants, and has long been
considered a candidate source for Galactic cosmic ray acceleration.  However,
the Larmor radius of 10 TeV protons in a \muG magnetic field 
is approximately 0.01\,pc, many orders of magnitude smaller than the distance to Vela, 
and unless unconventional propagation mechanisms are
assumed, charged particles from Vela will have lost all directional information
upon their arrival at Earth.  

Recently, several authors have investigated the extent to which the
stochastic nature of nearby supernova remnants can lead to spatial and temporal
variations in the cosmic ray flux \citep{Ptuskin20061909, Blasi:2011fm}.
The random nature of the sources makes quantitative predictions difficult, and
can lead to bumps and dips in the amplitude of the anisotropy as a function of
energy that depend on the specific source distribution used in the simulation
of the cosmic ray flux. Qualitatively, the models make specific
predictions for the energy dependence of the amplitude of the cosmic ray
anisotropy.

In the TeV-PeV range, the energy resolution of IceCube is poor for cosmic ray
events (see Sec.~\ref{subsec:dst}).  However, given the large rate of cosmic ray triggers, it is possible to
isolate a sufficiently large subset of showers with a median energy of several hundred
TeV which is not significantly contaminated by low energy events.  
A paper focusing on this study is currently in preparation.

The study of cosmic ray arrival directions at TeV energies will continue to be
a major ongoing research effort in IceCube.  IceCube and the future High
Altitude Water Cherenkov (HAWC) $\gamma$-ray observatory~\citep{Sinnis:2004je}
under construction in Mexico can be used to monitor the southern and northern
hemisphere, respectively, with high sensitivity.  The combined data sets will
soon allow for all-sky power spectra and the analysis of the entire sky at all
angular scales.

Over the next few years, with the IceCube detector now operating in its
complete 86-string configuration, our data set will increase at a rate of about $45 \times 10^9$
muon events per year. With this level of statistics we will also be able to study possible time
dependencies of the anisotropy in the southern hemisphere and compare to
similar studies performed with data from instruments in the northern
hemisphere~\citep{Abdo:2008aw,Amenomori:2010yr}.

\section{Acknowledgments}\label{sec:Acknowledgments}

We would like to thank Eric Hivon for helpful comments and suggestions
about the angular power spectrum analysis, and the Milagro Collaboration
for providing us with their data to produce the combined skymap
in Fig.\,\ref{fig:ic_plus_milagro}.

Some of the results in this paper have been derived using the
HEALPix~\citep{Gorski:2004by} and the
PolSpice~\citep{Szapudi:2000xj,Chon:2003gx} software libraries.

We acknowledge the support from the following agencies: U.S. National Science
Foundation-Office of Polar Programs, U.S. National Science Foundation-Physics
Division, University of Wisconsin Alumni Research Foundation, the Grid
Laboratory Of Wisconsin (GLOW) grid infrastructure at the University of
Wisconsin - Madison, the Open Science Grid (OSG) grid infrastructure; U.S.
Department of Energy, and National Energy Research Scientific Computing Center,
the Louisiana Optical Network Initiative (LONI) grid computing resources;
National Science and Engineering Research Council of Canada; Swedish Research
Council, Swedish Polar Research Secretariat, Swedish National Infrastructure
for Computing (SNIC), and Knut and Alice Wallenberg Foundation, Sweden; German
Ministry for Education and Research (BMBF), Deutsche Forschungsgemeinschaft
(DFG), Research Department of Plasmas with Complex Interactions (Bochum),
Germany; Fund for Scientific Research (FNRS-FWO), FWO Odysseus programme,
Flanders Institute to encourage scientific and technological research in
industry (IWT), Belgian Federal Science Policy Office (Belspo); University of
Oxford, United Kingdom; Marsden Fund, New Zealand; Japan Society for Promotion
of Science (JSPS); the Swiss National Science Foundation (SNSF), Switzerland;
A.~Gro{\ss} acknowledges support by the EU Marie Curie OIF Program;
J.~P.~Rodrigues acknowledges support by the Capes Foundation, Ministry of
Education of Brazil.

\end{document}